\documentclass[11pt, a4paper]{article}

\usepackage{accsupp}
\newcommand\numberstyle[1]{\tstyle\BeginAccSupp{ActualText={}}#1\protect\EndAccSupp{}\small}
\usepackage{listings} %
\usepackage[oldstyle]{hep-font} %
\usepackage{fullpage} %
\usepackage{hep-math-font} %
\usepackage{hep-math} %
\numberwithin{equation}{section}
\usepackage{hep-float} %
\graphics{figures}
\usepackage{hep-text} %
\usepackage{hep-bibliography} %
\bibliography{bibliography}
\usepackage{hep-reference} %
\usepackage{hep-acronym} %
\usepackage{hep-title} %
\usepackage[plot,feynman]{hep-graphic}
\graphicpath{graphics}
\tikzsetexternalprefix{tikz/}
\pgfplotsset{table/search path={data}}
\usetikzlibrary{positioning}
\usetikzlibrary{patterns.meta}
\usepgfplotslibrary{fillbetween}

\acronym{LNC}{lepton number conservation}
\acronym{LNV}{lepton number violation}
\acronymalternative{LNC}{lnc}{lepton number conserving}
\acronymalternative{LNV}{lnv}{lepton number violating}
\acronym{SM}{Standard Model}
\acronym{QFT}{quantum field theory}[quantum field theories]
\acronymalternative{QFT}{qft}{quantum field theoretical}
\acronym{EW}{electroweak}
\acronym{VEV}{vacuum expectation value}
\acronym{SPSS}{symmetry protected seesaw scenario}
\acronym{LHC}{Large Hadron Collider}
\acronym[HL-LHC]{HLLHC}{high luminosity phase of the \LHC}
\shortacronym{FCC}*{Future Circular Collider}
\shortacronym{ATLAS}{A Toroidal \LHC ApparatuS}
\shortacronym{CMS}{compact muon solenoid}
\acronym[LH$e$C]{LHeC}{Large Hadron Electron Collider}
\acronym{LNLS}{\emph{lepton number-like} symmetry}
\acronym{LO}{leading order}
\acronym{NLO}{next-to leading order}
\acronym{NDR}{no dispersion regime}
\acronym{TDR}{transverse dispersion regime}
\acronym{LDR}{longitudinal dispersion regime}
\acronym{MC}{Monte Carlo}
\acronym{JS}{Jacob-Sachs}
\acronym{BW}{Breit–Wigner}
\acronym{BM}{benchmark model}
\acronym{pSPSS}{phenomenological symmetry protected seesaw scenario}
\acronym[$N\mkern-1.5mu\widebar N$O]{NNO}{heavy neutrino-antineutrino oscillation}
\acronym[$0\nu$\beta\beta]{NLDB}{neutrinoless double \beta}
\acronym{EME}{energy-momentum envelope}
\acronym{STE}{spacetime envelope}

\renewcommand{\vec}{\mathbf}
\newcommand{\mat}{\mathsf}
\NewDocumentCommand{\expno}{om}{\IfNoValueF{#1}{#1\times}10^{#2}}
\newcommand{\expnumber}[2]{\expno[#1]{#2}}
\DeclareMathOperator*{\argmax}{argmax}

\renewcommand{\imaginaryunit}{i}
\renewcommand{\dot}{\cdot}
\newcommand{\proj}[1]{\widehat{#1}}
\newcommand{\identity}{\mathsftext 1}

\title{Decoherence effects on lepton number violation from heavy neutrino-antineutrino oscillations}

\author[basel]{Stefan Antusch}
\author[lisboa]{Jan Hajer}
\author[basel]{Johannes Rosskopp}

\affiliation[basel]{Departement Physik, Universität Basel, Klingelbergstrasse 82, CH-4056 Basel, Switzerland}
\affiliation[lisboa]{Centro de Física Teórica de Partículas (CFTP), Instituto Superior Técnico (IST), Universidade de Lisboa, 1049-001 Lisboa, Portugal}

\begin{document}

\maketitle

\begin{abstract}
We study decoherence effects and phase corrections in \NNOs, based on \QFTlong with external wave packets.
Decoherence damps the oscillation pattern, making it harder to resolve experimentally.
Additionally, it enhances \LNV for processes in symmetry-protected low-scale seesaw models by reducing the destructive interference between mass eigenstates.
We discuss a novel time-independent shift in the phase and derive formulae for calculating decoherence effects and the phase shift in the relevant regimes, which are the \NDRlong and \TDRlong.
We find that the phase shift can be neglected in the parameter region under consideration since it is small apart from parameter regions with large damping.
In the oscillation formulae, decoherence can be included by an effective damping parameter.
We discuss this parameter and present averaged results, which apply to simulations of \NNOs in the dilepton-dijet channel at the $\HLLHC$.
We show that including decoherence effects can dramatically change the theoretical prediction for the ratio of \lnv over $\lnc$ events.
\end{abstract}

\tableofcontents
\listoffigures
\listoftables
\lstlistoflistings

\clearpage

\section{Introduction}

The origin of the observed neutrino masses is one of the great open questions in current particle physics.
When the new particles involved in the neutrino mass generation have masses close to the \EW scale, it is possible to investigate this question at the \LHC and future accelerators.
One possible extension of the \SM of elementary particles that explains the observed light neutrino masses is based on the introduction of sterile neutrinos, \ie fermions which are uncharged under the gauge symmetry of the \SM \cite{Minkowski:1977sc}, see also \cite{Gell-Mann:1979vob,Mohapatra:1979ia,Glashow:1979nm,Yanagida:1980xy,Schechter:1980gr,Schechter:1981cv}.

When they form Yukawa interaction terms with the lepton and Higgs doublets and, in addition, have Majorana mass terms \cite{Majorana:1937vz}, light neutrino masses can be generated, which are then of Majorana-type.
However, when the sterile neutrino masses are around the \EW scale, care has to be taken not to exceed the bounds on the light neutrino masses \cite{Kersten:2007vk}.
When the Yukawa couplings are not tiny,  the smallness of the light neutrino masses is realised by an approximate \LNLS.
The sterile neutrinos then form pseudo-Dirac pairs of nearly mass-degenerate heavy neutrinos.
Although \LNV is significantly suppressed for prompt heavy neutrino decays, \cf \cite{Kersten:2007vk}, it can lead to observable effects via the phenomenon of \NNOs \cite{Antusch:2020pnn,Antusch:2022ceb}, see also \cite{Cvetic:2015ura,Anamiati:2016uxp,Antusch:2017ebe}.
Since the light neutrino masses become zero in the limit of exact \LNC, observing \lnv processes is crucial for probing the origin of neutrino masses.

Due to the \NNOs, the number of \lnc and \lnv events in a given process depends on the time difference between the production and decay of the heavy neutrinos.
Recently it has been shown for a selected \BM point consistent with present constraints, featuring a long-lived pseudo-Dirac heavy neutrino pair, that \NNOs could be resolved during the \HLLHC \cite{Antusch:2022hhh}, see also \cite{Antusch:2017ebe}.
However, even when the oscillations are not resolvable, they can induce \LNV.
The total ratio of \lnv over \lnc events, $R_{ll}$, can be used to quantify the effect.

Decoherence and phase correction effects on \NNOs are so far unexplored at the quantitative level.
Previous studies have used estimates to verify that decoherence effects can be neglected for the considered \BM parameters, \eg \cite{Antusch:2020pnn,Antusch:2022hhh}, or have assumed this to be the case.
While decoherence can, in principle, depend on various parameters, it has to be a function of the mass splitting of the heavy neutrinos.
This can be argued from the fact that for experimentally resolvable mass splittings, the pseudo-Dirac pair must reproduce the phenomenology of two separate Majorana neutrinos.
In such cases, \NNOs are expected to vanish.
Thus, in regions where decoherence effects are relevant, the simple \NNO formulae have to be modified.
Phase corrections for \NNOs have not yet been discussed.

One can calculate the possible decoherence and phase correction effects in \NNOs using \QFT with external wave packets.
The formalism is discussed in \cite{Beuthe:2001rc} and has been adapted to the case of \NNOs in \cite{Antusch:2020pnn}.
In \cite{Antusch:2022ceb}, the effective damping parameter $\lambda$ is introduced, which contains the collective effects of decoherence onto the \LO oscillation formulae.
In the present work, we explore how decoherence and a time-independent phase shift affect \NNOs as well as the quantitative prospects for observing \LNV.

The remainder of this publication is organised as follows:
In \cref{sec:external wave packets}, we introduce the external wave packet formalism.
Afterwards, in \cref{sec:damped oscillation probabilities}, we describe the derivation of the damped oscillation probability for the general case and subsequently apply the results to the case of \NNOs in the \SPSS.
We show that the effects of decoherence can be summarised by a damping parameter $\lambda$, leading to a simple extension of the oscillation formulae.
Results for $\lambda$, including its impact on $R_{ll}$ and searches for \LNV, are discussed in \cref{sec:results}.
Finally, we conclude in \cref{sec:conclusion}.
Details of the analytical derivations of the oscillation probabilities are presented in the appendices.
The detailed steps necessary to integrate the transition amplitude over the intermediate particles' momentum and travelled distance are presented in \cref{sec:momentum integration,sec:distance integration}, respectively.
The constant phase shift is discussed in \cref{sec:phase correction}.
The algorithm to compute the damping parameter $\lambda$ numerically is discussed in \cref{sec:pseudocode}, where the kinematics of the considered process is simulated using the \pSPSS introduced in \cite{Antusch:2022ceb,FR:pSPSS}.

\section{External wave packet formalism} \label{sec:external wave packets}

In this section, we derive the transition amplitude between two external states that are prepared as wave packets.
This essential quantity is the main ingredient to derive an oscillation probability following the arguments made in \cite{Antusch:2020pnn, Beuthe:2001rc}.
It is defined as a function of a distance $(t,\vec x)$ in spacetime
\footnote{Quantities with a suppressed vectorial index are indicated by boldface.}
\begin{equation} \label{eq:initial amplitude}
\mathcal A(t,\vec x) = \mel*{\Phi(t^{\prime\prime},\vec x^{\prime\prime})}{\mathcal T \exp\left[- \i \int[\d t^\backprime] \int[\d^3 \vec x^\backprime] \mathcal H(t^\backprime,\vec x^\backprime)\right] - \mathbb 1}{\Phi(t^\prime,\vec x^\prime)} \,,
\end{equation}
where $\mathcal H(t^\backprime,\vec x^\backprime)$ is the interaction Hamiltonian and $\mathcal T$ is the time ordering operator.
In comparison to the usual \qft approach, in which plane wave states $\ket{\Phi(\vec p)}$ with momentum $\vec p$ are used, the initial $\ket{\Phi(t^\prime,\vec x^\prime)}$ and final $\bra{\Phi(t^{\prime\prime},\vec x^{\prime\prime})}$ states are wave packets centred at the indicated points in spacetime and can be written as a function of a plane wave state using
\begin{equation}
\ket*{\Phi(t,\vec x)} = \int[\frac{\d^3 \vec p}{(2 \pi)^3 \sqrt{2 E(\vec p)}}] \psi(t,\vec x,\vec p,\vec p_0) \ket*{\Phi(\vec p)} \,,
\end{equation}
where  $E(\vec p)$ is the energy of the particle, and $\psi(t,\vec x,\vec p,\vec p_0)$ is the wave packet envelope which describes the shape of the wave packet and is centred around the momentum $\vec p_0$.
Assuming that the external wave packets are Gaussian and approximating the matrix element at the peak of those Gaussian functions, it is possible to evaluate the momentum integration over the external wave packets, yielding the transition amplitude for a mass eigenstate~$i$ \cite{Beuthe:2001rc}
\begin{equation} \label{eq:transition amplitude}
\mathcal A_i(t,\vec x) = \mathcal N \int[\d E] \int[\d^3 \vec p] M_i(E,\vec p) G_i(s) \exp\left[-f(E,\vec p) - \i \phi(t,\vec x,E,\vec p)\right] \,.
\end{equation}
Here $G_i(s)$ is the denominator of the renormalised propagator with $s = E^2 - \abs{\vec p}^2$, $M_i(E, \vec p)$ denotes the interaction amplitude, defined as the matrix element without the denominator of the propagator, and $\mathcal N$ is a normalisation constant.
\footnote{
The precise form of the normalisation constant $\mathcal N$ changes throughout the paper.
However, the normalisation constant can always be evaluated using an appropriate normalisation condition, as discussed in \cref{sec:NNO probability}.
}
The imaginary part of the exponent contains the phase
\begin{equation} \label{eq:phase}
\phi(t,\vec x,E,\vec p) = E t - \vec p \dot \vec x \,.
\end{equation}
Here $\vec x$ is the distance, and $t$ is the time difference between the production and detection point.
The real part of the exponent contains the \EME.
This name is derived from the fact that it describes the shape of the intermediate particle's wave packet as a function of its energy and momentum.
Its shape is defined by the shapes of the external particles' wave packets at the production $P$ and detection $D$ vertices $V$ and is given by \cite{Beuthe:2001rc}
\begin{align} \label{eq:EME}
f(E,\vec p) = \abs*{\frac{\vec p - \vec p_0}{2 \sigma_{\vec pP}^{}}}^2 + \left[\frac{e^{}_P(E, \vec p)}{2 \sigma_{EP}^{}}\right]^2 + (P\to D) \,,
\end{align}
where
\begin{align}
e^{}_V(E, \vec p) &= E - E_0 - (\vec p - \vec p_0) \dot \vec v_V \,, &
V &\in \{P, D\} \,,
\end{align}
Here $E_0$ and $\vec p_0$ are the energy and momentum of the intermediate particle obtained from the peaks of the external particles' wave packets using energy-momentum conservation either at the detection or production vertex.
They are thus called reconstructed energy and momentum.
The $(P\to D)$ is a shorthand notation where quantities at production are replaced by similar quantities at detection.
If the \EME is approximated as a Gaussian, its width can be interpreted as the \emph{effective width} $\sigma_\text{eff}$ of the intermediate particles' wave packet.

The total energy and momentum widths are given by the reciprocal sum of the respective widths at the production and detection vertices
\begin{align} \label{eq:total widths}
\frac1{\sigma_E^2} &= \frac{1}{\sigma_{EP}^2} + \frac{1}{\sigma_{ED}^2} \,, &
\frac1{\sigma_{\vec p}^2} &= \frac{1}{\sigma_{\vec pP}^2} + \frac{1}{\sigma_{\vec pD}^2} \,.
\end{align}
Each of these widths can be expressed in terms of the widths of the external particles in position space.
The widths of the external particles in position space are parameters of the theory and are determined by the experimental situation under consideration.
In the following, we only explicitly write definitions for quantities at production, while analogous definitions hold for quantities at detection.
The energy and momentum widths at this vertex are given by
\begin{align} \label{eq:energy-momentum width}
\frac{\sigma_{EP}^2}{\sigma_{\vec pP}^2} &= \Sigma_P - \abs{\vec v_P^{}}^2 \,, &
\sigma_{\vec pP}^{} \sigma_{\vec xP}^{} &= \frac{1}{2} \,, &
\frac{1}{\sigma_{\vec xP}^2} &= \sum_n \frac{1}{\sigma_{\vec xP_n}^2} \,,
\end{align}
where $\sigma_{\vec xP_n}^{}$ is the width of the external particle $n$ in position space and
\begin{equation}
\Sigma_P = \sigma_{\vec xP}^2 \sum_n \frac{\abs{\vec v_{P_n}^{}}^2}{\sigma_{\vec xP_n}^2} \,.
\end{equation}
The velocity of the production region is defined by
\begin{align} \label{eq:width velocities}
\vec v_P^{} &= \sigma_{\vec xP}^2 \sum_n \frac{\vec v_{P_n}^{}}{\sigma_{\vec xP_n}^2} \,, &
\vec v_{P_n}^{} &= \frac{\vec p_{P_n}^{}}{E_{P_n}} \,.
\end{align}
The particle with the smallest width dominates these terms unless its velocity is much smaller than the velocities of the other particles.
Since it holds that \cite{Beuthe:2001rc}
\begin{align}
0 &\leq \abs*{\vec v_P^{}}^2 \leq \Sigma_P \leq 1 \,, &
0 &\leq \Sigma_P - \abs{\vec v_P^{}}^2 \leq 1 \,,
\end{align}
one can calculate that the energy and momentum widths obey the inequality
\begin{equation} \label{eq:width estimate}
\sigma_E^{} \leq \sigma_{\vec p} \,.
\end{equation}

From the \EME \eqref{eq:EME}, it follows that energies $E$ and momenta $\vec p$ \emph{far} from the reconstructed energy $E_0$ and momentum $\vec p_0$ are exponentially suppressed, where \emph{far} is defined according to the energy or momentum width, respectively.
Additionally, damping from the propagator is expected when the reconstructed energy and momentum are such that the reconstructed mass
\begin{equation} \label{eq:reconstructed mass}
m_0^2 = E_0^2 - \abs{\vec p_0}^2
\end{equation}
is \emph{far} from the intermediate particles' masses $m_i$.
This damping defines the shape of the resonance, which in plane wave \QFT would be given by the Breit–Wigner distribution.

\section{Derivation of the damped oscillation probability} \label{sec:damped oscillation probabilities}

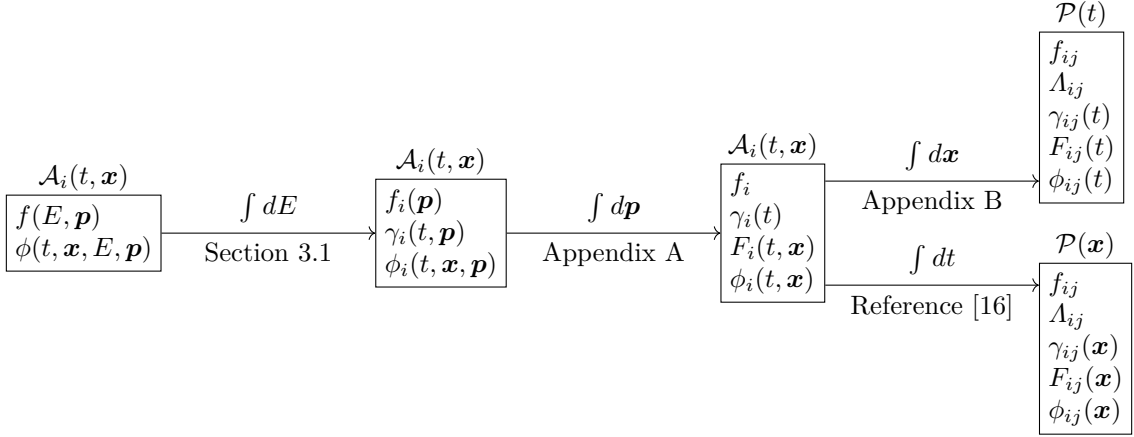
\begin{figure}
\begin{tikzpicture}[
 node distance=-3.5ex and 8em,
 header/.style={draw,align=left,label={[label distance=-.5ex]above:#1}},
 every path/.style={->},
]
\node(complete)[header={$\mathcal A_i(t,\vec x)$}]{$f(E,\vec p)$\\$\phi(t,\vec x,E,\vec p)$};
\node(energy)[right=of complete,header={$\mathcal A_i(t,\vec x)$}]{$f_i(\vec p)$\\$\gamma_i(t,\vec p)$\\$\phi_i(t,\vec x,\vec p)$};
\node(momentum)[right=of energy,header={$\mathcal A_i(t,\vec x)$}]{$f_i$\\$\gamma_i(t)$\\$F_i(t,\vec x)$\\$\phi_i(t,\vec x)$};
\node(length)[above right=of momentum,header={$\mathcal P(t)$}]{$f_{ij}$\\$\Lambda_{ij}$\\$\gamma_{ij}(t)$\\$F_{ij}(t)$\\$\phi_{ij}(t)$};
\node(time)[below right=of momentum,header={$\mathcal P(\vec x)$}]{$f_{ij}$\\$\Lambda_{ij}$\\$\gamma_{ij}(\vec x)$\\$F_{ij}(\vec x)$\\$\phi_{ij}(\vec x)$};
\draw(complete)--node[above]{$\int\d E$}node[below]{\Cref{sec:JS}}(energy);
\draw(energy)--node[above]{$\int\d\vec p$}node[below]{\Cref{sec:momentum integration}}(momentum);
\draw(momentum.45)--node[above]{$\int\d\vec x$}node[below]{\Cref{sec:distance integration}}(length.west|-momentum.45);
\draw(momentum.-45)--node[above]{$\int\d t$}node[below]{\Ccite{Beuthe:2001rc}}(time.west|-momentum.-45);
\end{tikzpicture}
\caption[Flowchart depicting the integration steps]{
\resetacronym{EME}%
Flowchart depicting the integration steps from the spacetime-dependent transition amplitude $\mathcal A_i(t,\vec x)$ of the mass eigenstate $i$ to the time and distance-dependent oscillation probabilities $\mathcal P(t)$ and $\mathcal P(\vec x)$, respectively.
The terms appearing in the exponential part of the transition amplitude and the oscillation probability are the \EME $f$, the phase $\phi$, the decay term~$\gamma$, the dispersion term $F$, and the localisation term $\Lambda$.
The last two terms give the main contribution to decoherence.
Thus they dominate the damping parameter $\lambda$.
} \label{fig:flowchart}
\end{figure}

The transition amplitude \eqref{eq:transition amplitude} needs to be integrated over the energy and momentum of the intermediate particle.
The strategy to perform these integration steps is depicted in \cref{fig:flowchart}.
While the energy integral is performed using the \JSlong theorem in the following section, the subsequent momentum integral is evaluated in \cref{sec:momentum integration}.
The final step is a distance average performed in \cref{sec:distance integration}, contrasting the time average used in \cite{Beuthe:2001rc}.
In this publication, we express the oscillation probability as a function of elapsed time instead of distance since the relevant observables naturally depend on the proper time of the oscillating particles rather than their travelled distance.

For example, since the heavy neutrinos, once they are detected at a collider experiment, will exhibit a range of Lorentz boosts, the oscillation pattern has to be translated into the proper time frame of the heavy neutrino in order to be reconstructable, see \cite{Antusch:2017ebe,Antusch:2020pnn,Antusch:2022hhh}.
Therefore, an oscillation probability as a function of time, averaged over the distance, is more suited for this purpose.
The same applies to measurements of the $R_{ll}$ ratio, which is sensitive to the interplay between \NNOs and the decay of heavy neutrinos and, hence, naturally defined in the proper time frame of the neutrinos.
Furthermore, a distance average is more straightforward from a technical point of view since there are fewer distance-dependent terms than time-dependent terms in the relevant exponential, as can be seen in \cref{fig:flowchart}.

\subsection{Energy integration via \JSlong theorem} \label{sec:JS}

\resetacronym{JS}

To further evaluate the transition amplitude \eqref{eq:transition amplitude}, the energy integral is evaluated using the \JS theorem \cite{Jacob:1961zz}.
This theorem states that for times larger than a threshold time~$t_{\JS}$, which is estimated in \cref{sec:JS estimation}, and for functions $\Psi(E, \vec p)$ that are non-zero only for a finite range of $s = E^2 - \abs{\vec p}^2$, the energy integral can be approximately evaluated using
\begin{equation}
\int[\d E] \Psi(E,\vec p) G_i(s) \exp\left[- \i E t\right] \approx \mathcal N \Psi(E_i^\prime(\vec p),\vec p) \exp\left[- \i E_i^\prime(\vec p) t\right] \,,
\end{equation}
where the complex pole energy is defined in terms of the complex pole of the propagator as
\begin{align} \label{eq:pole energy}
E_i^{\prime2}(\vec p) &= \abs{\vec p}^2 + z_i \,, &
z_i &= m_i^2 - \i m_i \Gamma_i \,.
\end{align}
while $m_i$ and $\Gamma_i$ are the mass and decay width of the mass eigenstate $i$, respectively.
After this approximate integration, the transition amplitude reads
\begin{equation} \label{eq:time integration amplitude}
\mathcal A_i(t,\vec x) = \mathcal N \int[\d^3 \vec p] M_i(E^\prime(\vec p),\vec p) \exp\left[- f(E_i^\prime(\vec p),\vec p) - \i \phi(t,\vec x,E_i^\prime(\vec p),\vec p)\right] \,.
\end{equation}
The pole energy can be rewritten as
\begin{equation} \label{eq:pole energy dependence}
E_i^{\prime2}(\vec p) =
\left[1 - 2 \i \epsilon_i(\vec p)\right] E_i^2(\vec p) \,,
\end{equation}
where the decay width expansion parameter, $\epsilon_i(\vec p)$, and the mass eigenstate energy, $E_i(\vec p)$, are defined as
\begin{align} \label{eq:decay width expansion parameter}
\epsilon_i(\vec p) &:= \frac{\gamma_i(\vec p)}{E_i(\vec p)} \,, &
\gamma_i(\vec p) &:= \frac{m_i \Gamma_i}{2 E_i(\vec p)} \,, &
E_i^2(\vec p) &= \abs{\vec p}^2 + m_i^2 \,.
\end{align}
Under the assumption that the decay width is small compared to the mass eigenstate energy, the phase, \eqref{eq:phase}, can be expanded in the decay width expansion parameter, which yields
\begin{align}
\phi(t,\vec x,E_i^\prime(\vec p),\vec p) &= \left[1 - \i \epsilon_i(\vec p) + \order*{\epsilon_i^2(\vec p)}\right] E_i(\vec p) t - \vec p \dot \vec x \,, &
\epsilon_i(\vec p) &\ll 1 \,.
\end{align}
The real part results in the phase of the mass eigenstate $i$, while the imaginary part generates an exponential decay term
\begin{align} \label{eq:time integrated phase and decay}
\phi_i(t,\vec x,\vec p) &:= E_i(\vec p) t - \vec p \dot \vec x \,, &
\gamma_i(t,\vec p) &:= \gamma_i(\vec p) t \,.
\end{align}
After the energy integration, the \EME \eqref{eq:EME} of the mass eigenstate $i$ can be approximated to be
\begin{equation}
f(E_i^{\prime}(\vec p),\vec p) = f(E_i(\vec p),\vec p) + \order*{\epsilon_i(\vec p)} \,,
\end{equation}
such that the \LO term reads
\begin{equation} \label{eq:time integrated EME}
f_i(\vec p) := f(E_i(\vec p),\vec p) = \abs*{\frac{\vec p - \vec p_0}{2 \sigma_{\vec pP}^{}}}^2 + \left[\frac{e^{}_{iP}(\vec p)}{2 \sigma_{EP}^{}}\right]^2 + (P\to D) \,,
\end{equation}
where
\begin{equation}
e^{}_{iV}(\vec p) := e^{}_V(E_i(\vec p), \vec p) = E_i(\vec p) - E_0 - (\vec p - \vec p_0) \dot \vec v_V \,.
\end{equation}
The derivation from which follows that higher orders in the decay width expansion parameter can generically be neglected is presented in \cref{sec:ND expansion}.
However, for very short times, the $\order*{\epsilon_i(\vec p)}$ terms can lead to
a time-independent phase shift, discussed in \cref{sec:phase correction}.
In the numerical calculation presented in \cref{sec:pseudocode}, these corrections are explicitly taken into account by identifying the imaginary part as a correction to the phase.

Finally, the transition amplitude \eqref{eq:transition amplitude} after the energy integration \eqref{eq:time integration amplitude} takes the form
\begin{equation} \label{eq:time integrated amplitude}
\mathcal A_i(t,\vec x) = \mathcal N \int[\d^3 \vec p] M_i(\vec p) \exp\left[-f_i(\vec p) - \gamma_i(t,\vec p) - \i \phi_i(t,\vec x,\vec p)\right] \,,
\end{equation}
where \NLO terms in the decay width expansion of the interaction amplitude $M_i(\vec p) = M_i(E_i,\vec p)$ are neglected.
The remaining integrals are the three-momentum integral and an integral that averages over distance or time.

\subsection{Applicability of the formalism} \label{sec:JS estimation}

\resetacronym{JS}

The \JS theorem used in the previous section is only valid for times larger than the \JSlong threshold time $t_{\JS}$.
It is defined via the diameter of the support of the intermediate particle's wave packet
\begin{equation}
t \geq t_{\JS} := \frac{1}{\abs*{\operatorname{supp}\exp\left[-f(E,\vec p)\right]}} \,.
\end{equation}
Since the interaction amplitude, and therefore the wave packet envelope, must vanish for values of $\sqrt s - m_i$ larger than the uncertainties, this support is estimated by the experimental uncertainty in reconstructing the mass $m_i$ in \cite{Beuthe:2001rc,Jacob:1961zz}.
In the case of \NNOs, where the mass of the heavy neutrino has to be reconstructed from semi-leptonic decay products, assuming this uncertainty to be of the order of one per cent of the heavy neutrino mass yields a time threshold of
\begin{equation}
t_{\JS} \approx \frac{100}{m} = \frac{\unit[1]{GeV}}{m} \unit[\expnumber{6.58}{-23}]{s} \,.
\end{equation}
In order to have a fraction $f$ of particles decaying beyond that time requires decay widths of
\begin{equation}
\Gamma \leq \Gamma_{\JS} := \frac{\gamma}{t_{\JS}} \ln \frac1f  \,,
\end{equation}
where $\gamma$ denotes the Lorentz boost factor.
Demanding \unit[99]{\%} of all particles decaying later than that time results in a \JS width of
\begin{equation}
\Gamma_{\JS} \approx \frac{m}{\unit[1]{GeV}} \unit[100]{MeV} \,,
\end{equation}
when assuming a Lorentz boost factor of $\gamma \approx 10$, which is a reasonable estimate for the parameter region considered in this work.

In contrast, the width of the wave packets of the external particles is estimated in \cref{sec:results} using the size of the silicon atom radius and the proton-proton distance in a beam bunch for final and initial states, respectively.
This line of argument suggests that the neutrino wave packet should be zero outside a range defined by the width of the wave packet in the squared four-momentum $s = p^2$
\begin{equation}
\frac{1}{2\sigma_s^2} = \frac{s}{2 E_0^2 \sigma_E^2} + \order*{\vec p - \vec p_0}
\end{equation}
Using the approximations derived in \cref{sec:wave packet widths dependencies}, the numerical values given in \cref{tab:widths} and further approximating $\sqrt s = m$, this estimate leads to a time threshold of
\begin{equation}
t_{\JS} \approx \frac{\gamma}{n \sigma_E^{}} \approx \frac{2 \gamma \sigma_p}{n} \approx \frac{\gamma}{n} \unit[200]{nm} \approx \frac{\gamma}{n} \unit[\expnumber{6.66}{-16}]{s}
\end{equation}
where $n$ is the number of standard deviations which is taken to define the support of the Gaussian distribution.
Requiring a fraction $f = \unit[99]{\%}$ of particles decaying later than that time leads to a decay width of
\begin{equation}
\Gamma_{\JS} \approx n \unit[0.0199]{eV} \,.
\end{equation}
Taking the \unit[5]{\sigma} range leads to a \JS decay width of about \unit[0.1]{eV}.
In order to be conservative, we use this more restrictive value in the following.

\subsection{Dispersion regimes of the momentum integration} \label{sec:dispersion regimes}

The integration over the three-momentum is carried out differently in three separate regimes depending on how fast the phase varies over the effective width.
For slowly varying phases, the integral is evaluated using Laplace's method.
This regime is called the \NDR since time-dependent dispersion effects can be neglected.
In this regime, the argument of the exponential in the transition amplitude \eqref{eq:time integrated amplitude} is expanded up to second order in the momentum $\vec p$ around the position of the minimum of the \EME at $\vec p_i$.
Therefore, the momentum $\vec p_i$ maximises the exponential of the \EME
\begin{equation}
\vec p_i = \argmax_{\vec p}\exp\left[-f_i(\vec p)\right] \,,
\end{equation}
The Hessian of the \EME \eqref{eq:time integrated EME} with respect to the momentum is given at \LO, \ie neglecting the mass splitting, by \eqref{eq:NDR EME Hessian}
\footnote{Quantities with suppressed matrix indices are indicated by sans-serif font.}
\begin{align} \label{eq:NDR inverse width}
\mat \Sigma_0 &= \frac{\identity}{2 \sigma_{\vec pP}^2} + \frac{\vec u_P^{} \otimes \vec u_P^{}}{2 \sigma_{EP}^2} + \left(P\to D\right) \,, &
\vec u_V^{} &:= \vec v_V^{} - \vec v_0 \,,
\end{align}
and defines the inverse of the \emph{effective width} of the intermediate particle.
\begin{equation} \label{eq:NDR effective width}
2 \mat \sigma_\text{eff}^2 = \mat \Sigma_0^{-1} \,.
\end{equation}
The matrix structure of $\mat \Sigma_0$ is defined by the two vectors $\vec u_P^{}$ and $\vec u_D^{}$.
Therefore, there exists a vector which is orthogonal to both, and the corresponding eigenvalue is
\begin{equation} \label{eq:smallest eigenvalue}
\abs*{\mat \Sigma_0}_\text{smallest} = \frac{1}{2 \sigma_{\vec p}^2} \,.
\end{equation}
Due to the inequality \eqref{eq:width estimate}, this is the smallest eigenvalue leading to the largest effective width.
The other two eigenvalues, which are dominated by the energy width, are therefore larger and approximately given by
\begin{equation} \label{eq:LO effective width}
\abs*{\mat \Sigma_0}_\text{larger} = \frac{\abs{\vec u}^2}{2\sigma_E^2} + \order*{\frac{\sigma_E^{}}{\sigma_{\vec p}}, \abs*{\vec v_P^{} - \vec v_D^{}}} \,.
\end{equation}
This approximation is justified when the velocity vectors are almost aligned $\vec u :\approx \vec u_P^{} \approx \vec u_D^{}$ and the inequality \eqref{eq:width estimate} is large $\sigma_E^{} \ll \sigma_{\vec p}$.
The \NDR applies to times shorter than the short-time threshold \eqref{eq:NDR short-time threshold}
\begin{align} \label{eq:short-time threshold}
t &\lessapprox t^\text{short} \,, &
t^\text{short} &= \abs{\mat \Sigma_0}_\text{smallest} E_0 = \frac{E_0}{2 \sigma_{\vec p}^2} \,.
\end{align}
Since, in its derivation, the phase is required to vary slowly over the effective width of the \EME in all directions, the short-time threshold depends on the largest effective width and, therefore, the smallest eigenvalue of the Hessian.
The detailed computation is described in \cref{sec:ND momentum integration}.

When wave packets travel longer, the phase oscillates more rapidly as a function of $\vec p$, such that Laplace's method, used in the short time regime, becomes unsuitable.
Since the wave packets are broader in directions transversal to the reconstructed momentum $\vec p_0$, an intermediate regime exists in which Laplace's method can only be used for the longitudinal direction.
In contrast, transversal directions are integrated using the method of stationary phase.
This intermediate regime is called the \TDR.
The method of stationary phase yields $p_x = p_y = 0$, assuming that the longitudinal component, indicated by hatted variables, is $\vec p_0 = \proj p_0 \vec e_z$.
Laplace's method in the longitudinal direction results in
\begin{align}
\eval{\vec p_i}_z &= \proj p_i = \argmax_{\proj p}\exp\left[-f_i(\proj p)\right] \,, &
\eval{\vec p}_x &= \eval{\vec p}_y = 0 \,.
\end{align}
The Hessian at \LO is given by \eqref{eq:TDR EME Hessian}
\begin{align} \label{eq:TDR inverse width}
\proj \Sigma_0 &= \frac{1}{2\sigma_{\vec pP}^2} + \frac{\proj u_P^2}{2 \sigma_{EP}^2} + (P\to D) = \frac{\proj u^2}{2 \sigma_E^2} + \order*{\frac{\sigma_E^{}}{\sigma_{\vec p}}, \abs{\proj v_P^{} - \proj v_D^{}}} \,, &
\proj u_V^{} &:= \proj v_V^{} - \proj v_0 \,,
\end{align}
where the last approximation holds for $\proj u :\approx \proj u_D^{} \approx \proj u_P^{}$ and when the inequality \eqref{eq:width estimate} is large,  \ie $\sigma_E^{} \ll \sigma_{\vec p}$.
Similar to the definition in the \NDR \eqref{eq:NDR effective width}, the effective width in the \TDR is defined as
\begin{equation} \label{eq:TDR effective width}
2 \proj \sigma_\text{eff}^2 = \proj \Sigma_0^{-1} \,.
\end{equation}
and the long-time threshold, which forms the upper bound for this regime \eqref{eq:TDR long-time threshold}, is defined by
\begin{align} \label{eq:long-time threshold}
t^\text{short} \lessapprox t &\lessapprox t^\text{long} \,, &
t^\text{long} &= \proj \Sigma_0 \frac{E_0^3}{m_0^2} = \frac{\proj u^2}{2 \sigma_E^2} \frac{E_0^3}{m_0^2} + \order*{\frac{\sigma_E^{}}{\sigma_{\vec p}}, \abs{\proj v_P^{} - \proj v_D^{}}} \,.
\end{align}
The computation leading to this result is presented in detail in \cref{sec:TD momentum integration}.

Longer times $t \gtrapprox t^\text{long}$ are not relevant for the discussion of heavy neutrinos in the parameter space of interest for this paper.
However, for the respective regime, called the \LDR, the distance-dependent formulae derived in \cite{Antusch:2020pnn,Beuthe:2001rc} can be used.

\begin{figure}
\begin{panels}{2}
\includetikz{Tdisp}
\caption{Lab frame.} \label{fig:dispersion regimes lab frame}
\panel
\includetikz{taudisp}
\caption{Proper time frame.} \label{fig:dispersion regimes proper time}
\end{panels}
\caption[Partition of the parameter space into regimes]{
\resetacronym{NDR}\resetacronym{TDR}\resetacronym{LDR}%
Partition of the parameter space into \LDR, \TDR, and \NDR through the long- and short-time thresholds as a function of the heavy neutrino mass $m$.
Panel \subref{fig:dispersion regimes lab frame} shows the partition in the lab frame distance $ct$ as defined in \eqref{eq:long-time threshold,eq:short-time threshold} and panel \subref{fig:dispersion regimes proper time} shows the partition in the proper time $\tau$ frame as defined in \eqref{eq:proper time thresholds}.
} \label{fig:dispersion regimes}
\end{figure}

The short- and long-time thresholds \eqref{eq:short-time threshold,eq:long-time threshold} are given in the lab frame.
Using $\tau = t \flatfrac{m_0}{E_0}$ they can be reexpressed in the proper time frame as
\begin{align} \label{eq:proper time thresholds}
\tau_\text{short} &= \abs{\mat \Sigma_0}_\text{smallest} m_0 = \frac{m_0}{2 \sigma_{\vec p}^2} \,, &
\tau_\text{long} &= \proj \Sigma_0 \frac{E_0^2}{m_0} = \frac{\proj u^2}{2 \sigma_E^2} \frac{E_0^2}{m_0} + \order*{\frac{\sigma_E^{}}{\sigma_{\vec p}}, \abs{\proj v_P^{} - \proj v_D^{}}} \,.
\end{align}

For heavy neutrinos appearing in the process presented in \cref{fig:feynman diagram}, these regimes are depicted in \cref{fig:dispersion regimes} after averaging over $100$ events.
The partition based on distances is presented in \cref{fig:dispersion regimes lab frame}.
It shows that for experimental length scales smaller than about \unit[100]{km}, only the \NDR and \TDR are relevant, and the short-time threshold is of $\order{\unit{dm}}$.
The regimes in the proper time frame are shown in \cref{fig:dispersion regimes proper time}.
For decay widths leading to lifetimes comparable to the short-time threshold, it becomes relevant to quantify the fraction of events that fall into the \NDR and the \TDR, respectively.
Decay widths $\Gamma \approx \tau_\text{short}^{-1}$ result in a fraction of $1-e^{-1}$ events decaying before the threshold, and therefore inside the \NDR.
For decay widths $\Gamma \gtrapprox 10 \tau_\text{short}^{-1}$ practically all events decay before the threshold.
Contrary, for decay widths $\Gamma \lessapprox 10^{-1} \tau_\text{short}^{-1}$, practically all events decay beyond the threshold, and therefore in the \TDR.
For decay widths $\Gamma \lessapprox \unit[10]{peV}$, the \LDR, becomes relevant.

\subsection{Time dependent oscillation probability}

The probability for a superposition of mass eigenstates $i$ and $j$ to yield the transition between the given initial and final states, defined in the amplitude \eqref{eq:initial amplitude}, is given by
\begin{equation} \label{eq:oscillation probability definition}
\mathcal P(t) = \mathcal N \int[\d \vec x]_{\vec x_0 - \Delta \vec x}^{\vec x_0 + \Delta \vec x} \sum_{ij} \mathcal A_i(t,\vec x) \mathcal A_j^*(t,\vec x) \,.
\end{equation}
The normalisation constant $\mathcal N$ can be evaluated using the condition
\begin{equation} \label{eq:normalisation condition}
\sum_{\mathclap{\text{outgoing}}} \mathcal P(t) = 1 \,,
\end{equation}
where the sum is understood to include all possible processes, \ie decay channels of the intermediate particle.
Since mass eigenstates acquire a complex phase while propagating, the superposition of distinct eigenstates depends on a phase difference that varies with time and distance, leading to a periodic fluctuation of the probability.
Therefore, we refer to the probability \eqref{eq:oscillation probability definition} as an oscillation probability.
The position space integral in this probability is performed in \cref{sec:ND distance integration} for the \NDR and in \cref{sec:TD distance integration} for the \TDR.
After this integration, the oscillation probability reads, according to results \eqref{eq:NDR oscillation probability,eq:TDR oscillation probability},
\begin{align} \label{eq:oscillation probability}
\mathcal P(t) &= \mathcal N \sum_{ij} M_{ij} \exp\left[- \lambda^\prime_{ij}(t) - \i \phi_{ij}(t)\right] \,, &
\lambda^\prime_{ij}(t) &= f_{ij} + \Lambda_{ij} + \gamma_{ij}(t) + F_{ij}(t) \,,
\end{align}
where the product of the interaction amplitudes with their momenta evaluated at the peak of the intermediate particle's wave packet is
\begin{align}
M_{ij} &= M_i M_j^* \,, &
M_i &= M_i(\vec p_i) \,.
\end{align}

From the definition of the oscillation probability \eqref{eq:oscillation probability definition}, it can be seen that it depends on the sum over the two mass eigenstates of the absolute value squared transition amplitudes.
Since the \EME \eqref{eq:time integrated EME} and the decay term \eqref{eq:time integrated phase and decay} are real-valued, the probability depends on their sum
\begin{align}
f_{ij} &= f_i + f_j \,, &
\gamma_{ij}(t) &= \gamma_i(t) + \gamma_j(t) \,,
\end{align}
where their values at the minimum of the \EME are given by
\begin{align}
f_i &= f(E_i, \vec p_i) \,, &
\gamma_i(t) &:= \gamma_i t \,, &
\gamma_i &:= \gamma_i(\vec p_i)  = \frac{m_i \Gamma_i}{2 E_i} \,, &
E_i^2 &= \abs{\vec p_i}^2 + m_i^2 \,.
\end{align}

Contrary, the exponential term describing the phase \eqref{eq:time integrated phase and decay} is imaginary, such that the probability depends on the phase difference calculated in \eqref{eq:NDR phase difference,eq:TDR phase difference},
\begin{align} \label{eq:oscillation}
\phi_{ij}(t) &= m_{ij} \tau(t) \,, &
m_{ij} &= m_i - m_j \,, &
\tau(t) &= \frac{m_0}{E_0} t \,,
\end{align}
where $\flatfrac{E_0}{m_0}$ is the Lorentz factor of the intermediate particle and $\tau(t)$ denotes the proper time the intermediate particle travels between production and decay.
It vanishes when the mass splitting becomes zero, \eg if $i = j$.
This expression is modified by subdominant \NLO terms \cite{Antusch:2020pnn,Beuthe:2001rc} and augmented by a time-independent shift, see \cref{sec:phase correction}.

From the derivation of the localisation term $\Lambda_{ij}$ in \eqref{eq:NDR localisation,eq:TDR localisation} as well as the dispersion term $F_{ij}(t)$ in \eqref{eq:TDR dispersion term}, it can be seen that they inherit this dependence on the mass splitting and that they are given by
\begin{align} \label{eq:decoherence}
\Lambda_{ij} &= \frac14
\begin{cases}
\vec p_{ij}^\trans \mat \Sigma_0 \vec p_{ij} & \NDR \,, \\
\proj \Sigma_0 \proj p_{ij}^2 & \TDR \,,
\end{cases} &
F_{ij}(t) &= \frac14
\begin{cases}
0 & \NDR \,, \\
\proj \Sigma_0^{-1} \proj v_{ij}^2 t^2 & \TDR \,,
\end{cases}
\end{align}
with
\begin{align}
\vec p_{ij} &= \vec p_i - \vec p_j \,, &
\proj p_{ij} &= \proj p_i - \proj p_j \,, &
\proj v_{ij} &= \proj v_i - \proj v_j \,,
\end{align}
where the inverse of the effective width $\Sigma_0$ is defined in \eqref{eq:NDR inverse width,eq:TDR inverse width}.
Both the localisation and the dispersion term are decoherence terms and thus lead to a damping of the oscillations.
While the time-dependent dispersion term is absent in the \NDR, it becomes relevant in the \TDR.

\subsection{\sentence\NNOlong probability} \label{sec:NNO probability}

\resetacronym{SPSS}
\resetacronym{NNO}

\begin{figure}
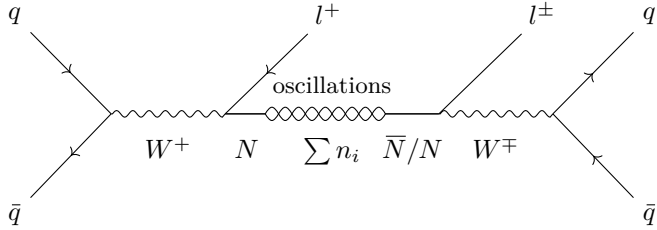

\includetikz*{oscillation-feynman}
\caption[Feynman diagram depicting production, oscillation, and decay]{
Feynman diagram depicting the heavy neutrino production at a hadron collider with subsequent oscillation and semi-leptonic decay.
In the external wave packet formalism, each external particle's width is a free parameter and needs to be adjusted according to the experimental setting.
We present our estimates for the external widths in \cref{tab:widths}.
} \label{fig:feynman diagram}
\end{figure}

From here on, we restrict to the \LO phenomenology of symmetry-protected low-scale seesaw models and \NNOs appearing in processes such as the one presented in \cref{fig:feynman diagram}.
The discussion is based on the \SPSS, recently introduced with its minimal phenomenological version, the \pSPSS, in \cite{Antusch:2022ceb}.
The generation of light neutrino masses in seesaw models is directly related to the presence of \LNV.
The process in \cref{fig:feynman diagram} is \lnc if the two charged leptons have opposite charges and \lnv if they have equal charge.
In this scenario the oscillation probability \eqref{eq:oscillation probability} for these two possible processes takes the form
\begin{equation} \label{eq:NNO probability}
\mathcal P^{\nicefrac{\LNV}{\LNC}}_{\alpha \beta}(t) = \mathcal N_\alpha(t) \sum_{i,j} V^{\nicefrac{\LNV}{\LNC}}_{\alpha \beta ij} \exp\left[- \lambda^\prime_{ij}(t) - \i \phi_{ij}(t)\right]\,.
\end{equation}
In comparison to reference \cite[][equation 2.28]{Antusch:2020pnn}, the exponential has been replaced by the one of the oscillation probability \eqref{eq:oscillation probability} containing, apart from the phase, additional terms due to the wave packet nature of the involved particles.
An additional term in \cite{Antusch:2020pnn}, which summarises the effects of the mass splitting in the interaction amplitudes, is neglected here since we treat the oscillations at \LO.
The factors of the leptonic mixing matrix at production $\alpha$ and decay $\beta$ are collected in the terms
\begin{align}
V^{\LNC}_{\alpha \beta ij} &:= V^{}_{\beta i} V^*_{\alpha i} V^*_{\beta j} V^{}_{\alpha j} \,, &
V^{\LNV}_{\alpha \beta ij} &:= V^*_{\beta i} V^*_{\alpha i} V^{}_{\beta j} V^{}_{\alpha j} \,.
\end{align}
For the \SPSS this results at \LO in \cite{Antusch:2022ceb}
\begin{equation}
V^{\nicefrac{\LNC}{\LNV}}_{\alpha \beta ij} =
\pm
\frac{\abs{\theta_\alpha}^2\abs{\theta_\beta}^2}4 \quad
\begin{cases}
\text{for \LNC and for \LNV with $i = j$} \,, \\
\text{for \LNV with $i\neq j$} \,.
\end{cases}
\end{equation}
where the active-sterile mixing angle is defined by
\begin{equation} \label{eq:mixing parameter}
\vec \theta = \vec y \frac{v}{m_M^{}} \,,
\end{equation}
with the \SM Higgs \VEV $v \approx \unit[174]{GeV}$ and the Yukawa coupling of one sterile neutrino labelled $\vec y$, see \cite{Antusch:2022ceb}.
The normalisation condition \eqref{eq:normalisation condition} for this scenario is evaluated for each flavour at production and yields
\begin{equation}
\label{eq:normalisation}
\begin{split}
1
= \sum_\beta \sum_{\substack{\LNC\\\LNV}} \mathcal P^{\nicefrac{\LNC}{\LNV}}_{\alpha\beta}(t)
&= \sum_\beta \sum_{\substack{\LNC\\\LNV}} \mathcal N_\alpha(t) \sum_{ij} V^{\nicefrac{\LNC}{\LNV}}_{\alpha \beta ij} \exp\left[- \lambda^\prime_{ij}(t) - \i \phi_{ij}(t)\right] \\
&= \sum_\beta \mathcal N_\alpha(t) \frac{\abs{\theta_\alpha}^2\abs{\theta_\beta}^2}{2} \sum_{i=j} \exp\left[- f_{ij} - \gamma_{ij}(t) \right] \,.
\end{split}
\end{equation}
In the last step, it has been used that the sum of leptonic mixing matrix factors over \lnc and \lnv processes vanishes for $i\neq j$.
Since the dispersion term, the localisation term, and the phase difference vanish for $i = j$, they are absent in the last line.

The oscillation probability \eqref{eq:NNO probability} between the two mass eigenstates $N_4$ and $N_5$ is then given by
\begin{align} \label{eq:LO NNO probability}
\mathcal P^{\nicefrac{\LNC}{\LNV}}_{\alpha\beta}(t) &= \frac{\abs{\theta_\beta}^2}{2 \sum_\gamma \abs{\theta_\gamma}^2} \left(1 \pm \exp\left[- \lambda_{45}(t)\right] \cos\left[\phi_{45}(t)\right]\right)& \forall& \alpha \,,
\end{align}
where the damping parameter takes the form
\begin{equation}
\exp[-\lambda_{45}(t)] =
\frac{2 \exp\left[- \lambda^\prime_{45}(t)\right]}{\exp\left[- f_{44} - \gamma_{44}(t) \right] +\exp\left[- f_{55} - \gamma_{55}(t) \right]} \,,
\end{equation}
and can be expressed as
\begin{equation} \label{eq:damping definition}
\lambda_{45}(t) := \Lambda_{45} + F_{45}(t) - \ln\sech\left[f_4 - f_5 + \gamma_4(t) - \gamma_5(t)\right] \,.
\end{equation}
Here $\sech(x)$ denotes the hyperbolic secant function, which is equal to one at the origin, and decays exponentially for values $\abs{x}\gg1$.
For the parameter region and time scales considered in this work, it is justified to
assume that the two decay parameters are approximately equal such that
\begin{align} \label{eq:damping approximated 1}
\lambda_{45}(t) &= \Lambda_{45} + F_{45}(t) - \ln\sech\left[f_4 - f_5\right] + \order*{\varepsilon} \,, &
\varepsilon &= \abs{\gamma_4(t) - \gamma_5(t)} \,,
\end{align}

From the \EME \eqref{eq:time integrated EME}, it can be seen that its minimum goes to zero if $m_i = m_0$.
However, heavy neutrinos with distinct masses cannot have $m_4 = m_0$ and $m_5 = m_0$ simultaneously.
Therefore, we consider two more limiting cases:

On the one hand, in cases where the reconstructed mass is \emph{near} the mean of the heavy neutrino masses, with respect to the energy and momentum widths, the values of the \EMEs are approximately equal $f_4 \approx f_5$.
The normalisation then cancels these contributions, such that the damping factor becomes
\begin{align} \label{eq:damping simplified}
\lambda_{45}(t) &= \Lambda_{45} + F_{45}(t) + \order*{\varepsilon^2} \,, &
\order{\varepsilon} &= \order[\big]{\abs*{\gamma_4(t) - \gamma_5(t)}} = \order[\big]{\abs*{f_4 - f_5}} \,.
\end{align}

On the other hand, configurations in which either the \EMEs or the decay terms are significantly different between the two mass eigenstates can lead to a damping of the oscillations.
For example, for mass splittings much larger than the energy and momentum widths, the minima $f_4$ and $f_5$ are very different.
The result is that one of the mass eigenstates is favoured by the available energy and momentum of the process, such that each event is dominated by one of the two Majorana particles, and the phenomenology is that of a pair of Majorana neutrinos without \NNOs.
If, \eg, $\varepsilon = \abs{\gamma_4(t) - \gamma_5(t)} \ll 1$ but $f_4 \ll f_5$ the damping parameter is given by
\begin{equation}
\begin{split}
\lambda_{45}(t) &= \Lambda_{45} + F_{45}(t) - \ln \sech f_5 + \order*{\varepsilon}
= \Lambda_{45} + F_{45}(t) + f_5 + \ln\frac{1 + e^{-2 f_5}}{2} + \order*{\varepsilon} \,.
\end{split}
\end{equation}
This leads to significant damping if $f_5 \gg 1$.
A similar argument holds for $\gamma_4(t) \ll \gamma_5(t)$.
The interpretation, in this case, is that if one of the mass eigenstates decays much faster than the other, oscillations are significantly suppressed, and damping is large.

The reconstructed mass $m_0$ has to be \emph{near} to one of the pole masses $m_4$ or $m_5$ since otherwise, the whole process is suppressed.
This effect is similar to the resonant scattering, described by an $s$-channel process with an intermediate particle of mass $m$.
For cases in which $\abs{s - m} \gg \Gamma$, the process is suppressed compared to $\abs{s - m} \ll \Gamma$.
In the present case, $s$ is labelled $m_0$, and the width of the resonance is dominated by the energy and momentum widths.

While the definition of the damping parameter is derived in the context of \NNOs in the \SPSS, the presented strategies for its evaluation also apply to more general processes.

\section{Damped \NNOslong} \label{sec:results}

For the simulation of the damped oscillations discussed in the previous section, the parameters that can impact the damping are
\begin{itemize}
\item The masses of the heavy neutrinos.
\item The decay widths of the heavy neutrinos, correlated with the time the heavy neutrinos propagate between production and decay.
\item The momentum configuration of the external particles.
\item The wave packet widths of the external particles.
\end{itemize}
The masses of the heavy neutrinos can be described in terms of their mean mass and their mass splitting
\begin{align}
m &= \frac{m_4 + m_5}{2} \,, &
\Delta m &= m_{45} = m_5 - m_4 \,.
\end{align}
The considered process and the heavy neutrinos' mean mass restrict the external particles' momentum configuration.
Since the exact momentum configuration changes on an event-per-event basis, a general result is obtained by averaging the computed damping parameter over several events.
Realistic momentum configurations are generated using the general purpose \MC generator \software{MadGraph5\_aMC@NLO} \cite{Alwall:2011uj} together with the \software{FeynRules} \cite{Alloul:2013bka} implementation of the \pSPSS defined in \cite{Antusch:2022ceb,FR:pSPSS}.
The numerical computation, obtained using the algorithm presented in \cref{sec:pseudocode}, takes the decay time of the heavy neutrino into account and is accordingly performed either in the \NDR or in the \TDR.

\begin{table}
\begin{panels}{2}
\begin{tabular}{r*6c} \toprule
& \multicolumn{3}{c}{Production} & \multicolumn{3}{c}{Detection} \\ \midrule
\multirow{2}{*}[-.75ex]{Particle} & $\widebar q$ & $q$ & $l$ & $l$ & $q$ & $\widebar q$ \\ \cmidrule{2-3} \cmidrule{6-7}
& \multicolumn{2}{c}{$W$} & & & \multicolumn{2}{c}{$W$} \\ \cmidrule(r){2-4} \cmidrule(l){5-7}
Width & \multicolumn{2}{c}{$\sigma_p$} & $\sigma_l^{}$ & $\sigma_l^{}$ & \multicolumn{2}{c}{$\sigma_j$} \\
\bottomrule \end{tabular}
\caption{Classes.} \label{tab:width classes}
\panel
\begin{tabular}{ccc} \toprule
$\sigma_p$ & $\sigma_l^{}$ & $\sigma_j$ \\ \midrule
$100$ & $0.111$ & $1.11$ \\
\bottomrule \end{tabular}
\caption{Values in \unit{nm}.} \label{tab:widths values}
\end{panels}
\caption[External wave packet widths]{
Position space widths assumed for the external wave packets of the incoming and outgoing particles appearing in the process presented in \cref{fig:feynman diagram}.
For simplicity, we assume that the widths fall into three distinct classes.
The widths of the incoming particles $\sigma_p$ are estimated by the \LHC beam bunch proton-proton distance.
The widths of the outgoing leptons $\sigma_l^{}$ are estimated by the atom radius of the silicon in the detector, and the width of the outgoing quarks and $W$ bosons $\sigma_j$ are estimated to be ten times as large.
Panel \subref{tab:width classes} shows the different classes the widths fall into, and panel \subref{tab:widths values} shows our baseline estimates assumed for the simulations performed in this work.
} \label{tab:widths}
\end{table}

In order to simulate the process shown in \cref{fig:feynman diagram}, the widths of the external particles' wave packets in position space need to be estimated.
When the heavy neutrino is lighter than the $W$ boson, the first $W$ boson can be on-shell such that its width can be directly estimated.
In contrast, the second $W$ boson is off-shell, and the external widths of its decay products must be estimated.
The situation is reversed if the heavy neutrino is heavier than the $W$ boson.
The wave packet widths in configuration space of the incoming particles $\sigma_p$ are assumed to be defined by the average distance between two protons in a beam bunch \cite{Apollinari:2017lan}.
The wave packet width of the outgoing leptons $\sigma_l^{}$ is assumed to be defined by the atom radius of silicon present in the detector material.
Final quarks and the final $W$ boson are expected to have a larger uncertainty than final leptons, such that their width $\sigma_j$ is given by $10 \sigma_l^{}$.
See \cref{tab:widths} for more details.

\subsection{Decay width dependence of the damping parameter} \label{sec:Decay width dependence of the damping parameter}

\begin{figure}
\begin{panels}{2}
\includetikz{fixedmass}
\caption{Onset of damping.} \label{fig:damping fixed m num}
\panel
\includetikz{fixedmass-2}
\caption{Validity of approximations.} \label{fig:damping fixed m ana}
\end{panels}
\caption[Damping parameter $\lambda$ as a function of $\Gamma$ and $\Delta m$.]{
\resetacronym{NDR}\resetacronym{TDR}\resetacronym{LDR}%
Damping parameter $\lambda$ as a function of $\Gamma$ and $\Delta m$.
The contour lines for the damping parameter are shown for two different masses $m$ as a function of the decay width $\Gamma$ and the mass splitting $\Delta m$.
The numerical results for the damping parameter $\lambda$ averaged over $500$ events per parameter point are given four fixed values of $\exp(-\lambda)$ and in panel \subref{fig:damping fixed m num}.
The comparison between the numerical results and the analytical approximations for the \NDR and the \TDR are given for $\lambda = \ln 2$ in panel \subref{fig:damping fixed m ana}.
The mass dependence is given in more detail in \cref{fig:damping}.
Very small decay widths of $\Gamma \lessapprox \unit[10]{peV}$ are governed by the \LDR, which is not calculated in this work.
Therefore, the predictions for these events are simulated using the same techniques as for the events falling into the \TDR and hence are not reliable.
Note that in the $m=\unit[500]{GeV}$ case, the analytical \TDR line and the numerical line coincide.
} \label{fig:damping fixed m}
\end{figure}

The mean decay width of the heavy neutrinos determines the time range the neutrino can propagate before it decays.
It is thus possible to examine the time dependence of the damping parameter $\lambda = \lambda_{45}(t)$ by studying its dependence on the decay width.
A numerical computation of the damping parameters for fixed mean masses of heavy neutrinos is presented in \cref{fig:damping fixed m num}.
The shape of the contours depicting constant damping consists of three regions:
\begin{itemize}

\item To the right is a plateau stretching over several orders of magnitude.
The plateau demonstrates that the effects due to varying decay widths, and with it, the time dependence of $\lambda$, are not significant in this region.
The plateau can be understood from the result \eqref{eq:decoherence}, noting that neither the momentum differences $\vec p_{ij}$ nor the matrix $\Lambda_{ij}$ contains any terms in $\Gamma$ or $t$ at \LO.

\item For smaller decay widths $\Gamma \lessapprox \unit[0.1]{\mu eV}$ the damping increases with decreasing decay width.
This effect is due to non-identical group velocities of wave packets of different mass eigenstates, which causes the wave packets to separate over time and, in turn, causes decoherence.
The effect becomes larger for heavy neutrinos that live longer.
However, if only the first $100$ oscillation cycles are considered, the effect vanishes, and the plateau in the central section continues for small decay widths.
Alternatively, the effect also vanishes if decays inside a sphere of radius \unit[50]{cm} are considered.
\footnote{
Note that the \unit[50]{cm} represents a somewhat randomly chosen value for which we have checked that the effects can be neglected.
It does not represent a boundary at which those effects become relevant.
}
Since these two restrictions cover most phenomenologically interesting cases, the effects of decoherence due to the separation of wave packets can be neglected in the parameter region under consideration.
In the following discussions, we assume these restrictions.
They imply that the damping parameter depends, in addition to the width of the external wave packet in position space, only on the mean mass and the mass splitting of the heavy neutrinos, \ie $\lambda = \lambda(m, \Delta m)$.

\end{itemize}

In \cref{fig:damping fixed m} the plateau extends beyond $\Gamma = \unit[0.1]{eV}$, until where our numerical calculations are applicable, as we discussed in \cref{sec:JS estimation}.
Since the physics leading to the damping as a function of $\Delta m$ is independent of $\Gamma$ at \LO according to our analytical derivations in \cref{sec:phase correction}, we conjecture that we can extrapolate the plateau also to larger $\Gamma$.
We make use of this conjecture when we analyse the consequence of damping on $R_{ll}$ in \cref{sec:LNV,sec:Rll}.

The analytical damping formula \eqref{eq:damping approximated 1}, together with the approximated expressions for decoherence \eqref{eq:decoherence}, reproduces the plateau found in the numerical evaluation for both regimes, as shown in \cref{fig:damping fixed m ana}.
Since the time-dependent dispersion is disregarded in the \NDR, the respective formulae do not feature the increased damping for small decay widths.
In the region where they are applicable, the analytical formulae are in good agreement with the numerical results for $m=\unit[10]{GeV}$ and in almost perfect agreement for $m=\unit[500]{GeV}$.

\subsection{Mass dependence of the damping parameter} \label{sec:Mass dependence of the damping parameter}

\begin{figure}
\begin{panels}{2}
\includetikz{mdeltamnum}
\caption{Numerical derivation.} \label{fig:damping numeric}
\panel
\includetikz{mdeltamT}
\caption{Time dependent formulae.} \label{fig:damping distance integrated}
\panel
\includetikz{mdeltamL}
\caption{Distance dependent formulae.} \label{fig:damping time integrated}
\panel
\includetikz{mdeltam}
\caption{Overall comparison.} \label{fig:damping comparison}
\end{panels}
\caption[Damping parameter $\lambda$ as a function of $m$ and $\Delta m$]{%
\resetacronym{NDR}\resetacronym{TDR}%
Simulation results for the damping parameter $\lambda$ for four values of $\exp(-\lambda)$ as a function of the mass $m$ and the mass splitting $\Delta m$ using the numerical analysis from \cref{sec:pseudocode} in panel \subref{fig:damping numeric}, the time-dependent results from \cref{sec:distance integration} in panel \subref{fig:damping distance integrated}, and the distance-dependent results from \cite{Beuthe:2001rc} in panel \subref{fig:damping time integrated}.
Panel \subref{fig:damping comparison} compares these techniques.
For the analytic results in panels \subref{fig:damping distance integrated} to \subref{fig:damping comparison}, the results for the \NDR and \TDR are shown separately.
} \label{fig:damping}
\end{figure}

With the restrictions established in the last section, the damping parameter is time-independent and can be studied as a function of the mean mass $m$ and the mass splitting $\Delta m$ of the heavy neutrino.
This dependency is presented in \cref{fig:damping}.
The numerical results are shown in \cref{fig:damping numeric}, the results for the time-dependent analytic formulae \eqref{eq:damping approximated 1} are presented in \cref{fig:damping distance integrated}, and the results for a distance-dependent oscillation probability, as obtained in \cite{Antusch:2020pnn,Beuthe:2001rc}, are shown in \cref{fig:damping time integrated}.
While the numerical results in the \NDR and the \TDR are approximately equal, the results for the approximated analytic formulae derived in \cref{sec:momentum integration,sec:distance integration} differ between those regimes.
Therefore, the results obtained from the analytical computation are presented for each regime individually, while the presented numerical results are valid in both regimes.

Since the time-dependent formulae for the damping factor are similar in the \NDR and \TDR, the \LO effects of the momentum dependence can be explained by studying the time-independent part of the damping factor \eqref{eq:damping simplified}
\begin{equation} \label{eq:damping approximated}
\lambda_{45} = \Lambda_{45} + \order*{\varepsilon, \frac{t}{t^\text{short}}} = 2\left(\vec p_{45} \dot \frac{\vec u_P^{}}{2 \sigma_{EP}^{}}\right)^2 + \order*{\varepsilon, \frac{t}{t^\text{short}}, \frac{\sigma_{\vec p}}{\sigma_E^{}}} \,,
\end{equation}
where \eqref{eq:LO effective width,eq:decoherence} are used, and the last approximation is obtained by observing that, in both regimes, the energy width at production dominates the reciprocal sum in the localisation term for the baseline estimate of external widths, defined in \cref{tab:widths values}.
Although the exact dependence of the damping factor on the mass is complicated since the process-dependent orientation of momenta and velocities change with varying mass, the sudden decrease of damping around the $W$ boson mass can be explained by a change in the energy width.
The energy width at production is given by a sum of all external particles at the production vertex.
For heavy neutrinos lighter than the $W$ mass, this includes the initial $W$ boson and the initial charged lepton.
For heavy neutrinos above the $W$ mass, the initial $W$ boson is off-shell, and thus, the relevant particles are given by the two incoming quarks and the initial charged lepton.
An increase in the number of external particles participating in the production process results in a sudden increase in the energy width, which results in a sudden decrease in damping.

The shape of the contours describing constant damping is very similar for all computation methods in all regimes, except for the distance-dependent formulae in the \NDR.
This difference can be traced back to the localisation term in the \NDR, which takes the form
\begin{equation} \label{eq:localisation time vs distance integrated}
\Lambda_{ij} = \frac12
\begin{Bmatrix}
\vec p_{ij}^\trans (\partial_{\vec x,\vec x} F_{ij}(t, \vec x))^{-1} \vec p_{ij} \\
E_{ij} (\partial_{t,t} F_{ij}(t, \vec x))^{-1} E_{ij}
\end{Bmatrix}
= \frac{1}{4}
\begin{cases}
\vec p_{ij} ^\trans \mat \Sigma_0^{} \vec p_{ij}  & \text{time dependent} \,, \\
E_{ij}^2\left(\vec v_0 ^\trans \mat \Sigma_0^{-1} \vec v_0\right)^{-1} & \text{distance dependent} \,.
\end{cases}
\end{equation}
where $E_{ij} := E_i - E_j$ and $F_{ij}(t, \vec x)$ is the \STE in the \NDR \eqref{eq:NDR STE LO}.
The time-dependent formula is given in \eqref{eq:decoherence}, and the distance-dependent one can be found in \cite[][section 6.1.2, equation 99]{Beuthe:2001rc}.
While the upper formula is proportional to the eigenvalues of $\mat \Sigma_0$, the lower formula is proportional to the inverse of the eigenvalues of $\mat \Sigma_0^{-1}$.
Therefore, as long as the reconstructed neutrino velocity $\vec v_0$ is not parallel to one of the eigenvectors of the effective width, this lead to significantly different damping behaviour.

The absence of the dispersion term in the \NDR can yield different damping for each regime.
However, for the restrictions discussed in \cref{sec:Decay width dependence of the damping parameter}, the effects of this term are expected to be negligible.
Together with the fact that all other additional effects considered in the numerical derivation compared to the analytic one are small, the numerical and time-dependent analytical results in both regimes are expected to be approximately identical.
This is confirmed by the results shown in \cref{fig:damping}.
Therefore, the time-dependent results feature a smooth transition of the damping between the \NDR and the \TDR.
In contrast, for the distance-dependent results, the damping in the \NDR differs significantly from the results in the \TDR.
This smooth transition can be seen as a further advantage of the time-dependent formulae, as derived in this work, over the distance-dependent ones.

\subsection{Wave packet widths dependencies} \label{sec:wave packet widths dependencies}

\begin{figure}
\begin{panels}{2}
\includetikz{taudisp2}
\caption{Time thresholds.} \label{fig:regime dependence}
\panel
\includetikz{mdeltam2}
\caption{Damping parameter.} \label{fig:damping dependence}
\end{panels}
\caption[Parameter dependence on external wave packet widths]{
Dependence of the time thresholds and the damping parameter on the scaling of each external wave packet width.
The values for the widths of the baseline scenarios are given in \cref{tab:widths}.
The baseline scenario for the time thresholds shown in panel \subref{fig:regime dependence} is given in \cref{fig:dispersion regimes}, and the baseline scenario for the damping parameter in panel \subref{fig:damping dependence} is given in \cref{fig:damping numeric}.
The short- and long-time thresholds in panel \subref{fig:regime dependence} are most sensitive to the scaling of the lepton $\sigma_l^{}$ and the proton $\sigma_p$ widths, respectively.
The damping parameter in panel \subref{fig:damping dependence} is most sensitive to the scaling of the proton width.
} \label{fig:observable dependence}
\end{figure}

In order to understand the impact of the estimates for the widths of the external wave packets, we rederive the previous results after rescaling individual widths by a factor of one hundred and present their dependence on this scaling in \cref{fig:observable dependence}.
\footnote{This scaling factor is chosen to generate large deviations from the baseline estimates to illustrate the parameter dependence and does not represent uncertainty in the baseline estimates.}

The effects on the relevant time thresholds are presented in \cref{fig:regime dependence}.
They can be understood by considering their dependence on the external width given in \eqref{eq:short-time threshold,eq:long-time threshold}.
While the momentum width $\sigma_{\vec p}$ depends on the smallest width in configuration space at production and detection, which is given by the widths of the charged leptons $\sigma_l^{}$, the energy width $\sigma_E^{}$ depends, according to the following argument, mainly on the proton width $\sigma_p$.

The energy width depends on the two smallest widths at production and detection.
While the smallest width is cancelled by the global factor of $\sigma_{\vec p P}$ or $\sigma_{\vec p D}$, respectively, the second smallest width dominates the energy width \eqref{eq:energy-momentum width}.
For the baseline of external widths, see \cref{tab:widths}, the energy width at production is much smaller than the corresponding energy width at detection.
The reciprocal sum of those energy widths, precisely $\sigma_E^{}$, is thus dominated by the proton width.

In \cref{fig:damping dependence}, we show that the main impact on the damping parameter is due to the proton width by individually varying the external widths.
This relation can be traced back to the fact that the damping is dominated by the energy width, see approximation \eqref{eq:damping approximated}.

The effect of the non-monotonous behaviour of the contour around the $W$ boson mass in \cref{fig:damping dependence} is reduced when varying the jet width.
The non-monotonous behaviour can be explained by a change in the number of particles participating in production and detection, as described in \cref{sec:Mass dependence of the damping parameter}.
The change in the number of particles is the opposite for production and detection, such that it weakens if the energy widths at production and detection are similar.
Multiplying the jet width by a factor of $100$ has precisely the effect of equalising those energy widths.

From the above results, it becomes clear that the value of the external widths plays an essential role in the prediction of decoherence and merits further dedicated studies.

\subsection{Decoherence effects on $R_{ll}$} \label{sec:Rll}

The ratio between the number of \lnv and \lnc decays is called $R_{ll}$.
Since it is calculated by integrating over the \NNOs, it is affected by decoherence.
Therefore, it is necessary to know the amount of damping to predict its value as a function of, \eg, the mean mass and the active-sterile mixing parameter.

The probability of obtaining an \lnc or \lnv event, between proper times $\tau_{\min}$ and $\tau_{\max}$, is given by the integral \cite{Antusch:2022ceb}
\begin{equation}
P_{ll}^{\nicefrac{\LNC}{\LNV}}(\tau_{\min}, \tau_{\max}) = \int_{\tau_{\min}}^{\tau_{\max}}
P_\text{decay}(\tau)
P^{\nicefrac{\LNC}{\LNV}}_\text{osc}(\tau)
\d \tau \,,
\end{equation}
where $\tau = \nicefrac{m_0}{E_0} t$ is the proper time.
Here,
the decay probability density is given by
\begin{equation}
P_\text{decay}(\tau) = - \dv \tau \exp\left(- \Gamma \tau\right) = \Gamma \exp\left(- \Gamma \tau\right) \,,
\end{equation}
and the oscillation probability is given by \eqref{eq:LO NNO probability}
\begin{equation}
P^{\nicefrac{\LNC}{\LNV}}_\text{osc}(\tau) = \mathcal N \left(1 \pm \exp\left[- \lambda - \frac{\mu^2 \tau^2}{4}\right] \cos\left[\Delta m \tau\right]\right) \,,
\end{equation}
where the time dependence of the damping parameter \eqref{eq:damping approximated 1} has been made explicit by defining the parameter $\mu$ using the dispersion term in the \TDR \eqref{eq:decoherence}.
In the limit $\tau_{\min} \rightarrow 0$ and $\tau_{\max}\rightarrow \infty$ the integral and the ratio of \lnv over \lnc events is given by
\begin{align}
P_{ll}^{\nicefrac{\LNC}{\LNV}}(\lambda,\mu) &\propto \frac{1 \pm f(\lambda,\mu)}{2} \,, &
R_{ll}(\lambda,\mu) &= \frac{2}{1 + f(\lambda, \mu)} - 1 \,.
\end{align}
where the function that appears in both quantities is
\begin{equation}
f(\lambda,\mu) = \frac{\operatorname{erfcx}[\Gamma^\prime_-(\mu)] + \operatorname{erfcx}[\Gamma^\prime_+(\mu)]}{2} \Gamma^\prime_\lambda(\mu) \,,
\end{equation}
which is defined in terms of
\begin{align}
\operatorname{erfcx}(x) &= \exp(x^2) \left[1 - \operatorname{erf}(x)\right] \,, &
\Gamma^\prime_\pm(\mu) &= \frac{\Gamma \pm \i \Delta m}{\mu} \,, &
\Gamma^\prime_\lambda(\mu) &= \frac{\Gamma}{\mu} \frac{\sqrt \pi}{\exp \lambda} \,.
\end{align}
Here $\operatorname{erfcx}(x)$ is the scaled complementary error function, which decays exponentially for negative $x$ approaches one for small $x$ and is inversely proportional to $x$ for large $x$.
For a subleading time dependence in the damping parameter, the function  can be approximated using
\begin{align}
f(\lambda,\mu) &= f(\lambda,0) \left(1 - \frac{\Gamma^2 - 3 \Delta m^2}{\Gamma^2 + \Delta m^2} \frac\varepsilon2 + \order*{\varepsilon^2}\right) \,, &
\varepsilon &= \frac{1}{\Gamma_-^\prime(\mu) \Gamma_+^\prime(\mu)} = \frac{\mu^2}{\Gamma^2 + \Delta m^2} \,.
\end{align}
The \LO term corresponds to the limit $\mu \to 0$, which captures the \NDR.
In this limit the equations simplify to
\begin{align} \label{eq:Rll of lambda}
f(\lambda,0) &= \frac{f(0,0)}{\exp \lambda} \,, &
R_{ll}(\lambda) &:= R_{ll}(\lambda,0) = 1 - \frac{2}{1 + \left(1 + \nicefrac{\Delta m^2}{\Gamma^2}\right) \exp \lambda} \,,
\end{align}
where the damping independent term $f(0,0)$ corresponds to the term that appears when taking furthermore the limit that also the constant localisation term can be neglected, $\lambda\to0$, which recovers for the ratio of \lnv over \lnc events the known result \cite{Anamiati:2016uxp,Antusch:2022ceb}
\begin{align} \label{eq:naive Rll}
f(0,0) &= \frac{\Gamma^2}{\Gamma^2 + \Delta m^2} \,, &
R_{ll}(0,0) &= \frac{\Delta m^2}{\Delta m^2 + 2 \Gamma^2} \,.
\end{align}

\begin{figure}
\begin{panels}{2}
\includetikz{Rllalpha}
\caption{$R_{ll}(\lambda)$} \label{fig:Rll of lambda}
\panel
\includetikz{Rll}
\caption{$R_{ll}(\Gamma, \Delta m)$} \label{fig:Rll of Gamma and Delta m}
\end{panels}
\caption[Decoherence effects on the $R_{ll}$ ratio]{
Decoherence effects on the $R_{ll}$ ratio for values in $R_{ll} \in [0.1,0.9]$.
Panel \subref{fig:Rll of lambda} depicts the functional dependence given in \eqref{eq:Rll of lambda} as a function of the naive value \eqref{eq:naive Rll}, \ie assuming $\lambda = 0$ and the damping parameter.
The simulation results for the impact of decoherence effects in the $(\Gamma,\Delta m)$ parameter space are presented in panel \subref{fig:Rll of Gamma and Delta m}.
Results for the damping parameter $\lambda$ are used, as discussed in \cref{sec:Decay width dependence of the damping parameter}.
} \label{fig:Rll}
\end{figure}

In cases where the damping is large $\lambda(\tau) \gg 1$, the coherence between the propagating mass eigenstates is lost, and the phenomenology is that of two independent Majorana neutrinos.
Increasing $\lambda$, therefore, increases the observed $R_{ll}$ compared to the naive case that does not take damping into account.
This behaviour is depicted for time-independent damping in \cref{fig:Rll of lambda}.
Therefore, when considering $R_{ll}(\lambda)$ as a function of $\Delta m$ and $\Gamma$, parameter regions with large damping must have a large $R_{ll}$ and lines representing a smaller $R_{ll}$ cannot penetrate those regions.
As depicted in \cref{fig:Rll of Gamma and Delta m}, the contours representing the naive $R_{ll}$ are given by constant ratios between $\Gamma$ and $\Delta m$.
However, damped $R_{ll}$ contours are bound from above by regions of large damping.
Therefore, once damping becomes relevant, the $R_{ll}$ bands follow the contour lines of the damping parameter shown in \cref{fig:damping fixed m}.

\subsection{Decoherence effects on prompt searches for \LNVlong} \label{sec:LNV}

\begin{table}
\begin{tabular}{lclr@{${}={}$}l} \toprule
Seesaw & $\Delta m$ & Hierarchy & \multicolumn{2}{c}{$\BM$} \\ \midrule
\multirow{2}{*}{Linear} & \multirow{2}{*}{$\Delta m_\nu$} & Normal & $\Delta m_\nu$ & $\unit[(41.46\pm0.29)]{meV}$ \\
 & & Inverted & $\Delta m_\nu$ & $\unit[(749\pm21)]{\mu eV}$ \\ \cmidrule{3-5}
\multirow{3}{*}{Inverse} & \multirow{3}{*}{$m_\nu \abs{\vec \theta}^{-2}$} & & $m_\nu$ & $\unit[0.5]{meV}$ \\
 & & & $m_\nu$ & $\unit[5]{meV}$ \\
 & & & $m_\nu$ & $\unit[50]{meV}$ \\
\bottomrule \end{tabular}
\caption[Benchmark models]{
\resetacronym{BM}%
Summary of the five \BMs considered in this publication, resulting in various mass splittings of the heavy neutrinos $\Delta m$.
Since the \SPSS with one pseudo-Dirac pair captures the minimal linear seesaw, it suffices to define one \BM for each light neutrino mass hierarchy that reproduces the observed mass splitting between two massive light neutrinos $\Delta m_\nu$ \cite{Esteban:2020cvm}.
In the case of the inverse seesaw, the single pseudo-Dirac \SPSS represents an incomplete theory since the model generates only one of the two light neutrino masses.
Therefore, we define the mass splitting as a function of the generated neutrino mass $m_\nu$, for which three \BM values are defined.
} \label{tab:BMs}
\end{table}

\begin{figure}
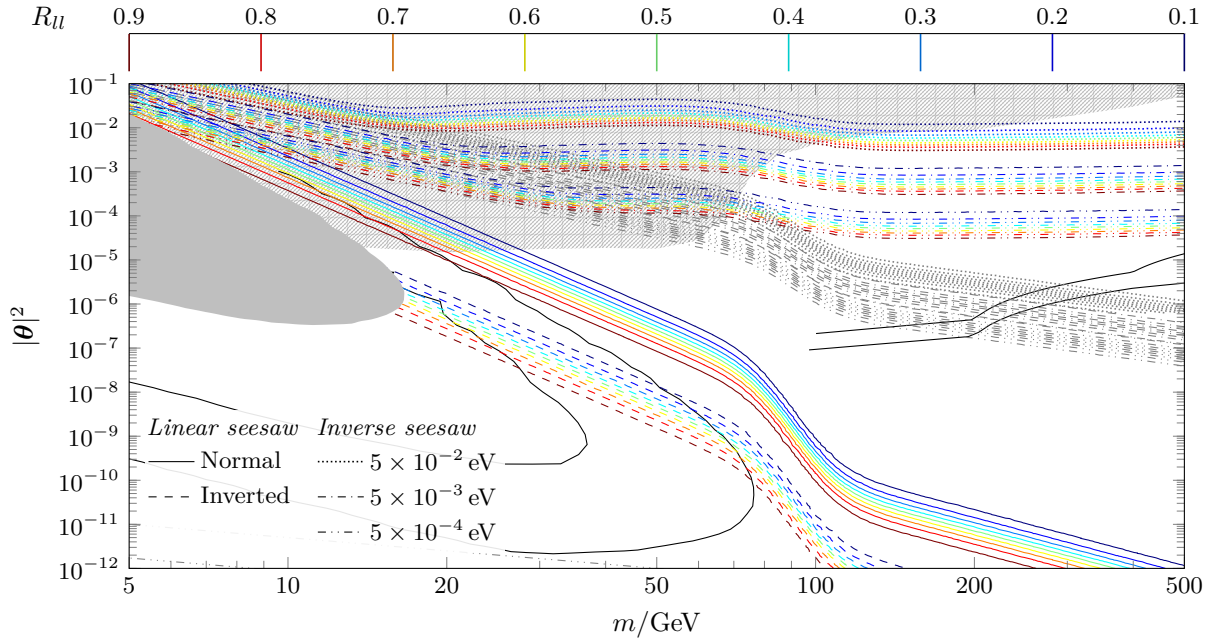

\includetikz{Rllmtheta}
\caption[Decoherence effects on prompt $\lnv$ searches]{
Bands of the $\lnv$ over $\lnc$ event ratio $R_{ll} \in [0.1,0.9]$ for the five $\BMs$ defined in \cref{tab:BMs}.
The genuine results, taking damping into account as discussed in \cref{sec:Decay width dependence of the damping parameter}, are depicted as coloured bands, while the naive results that neglect damping are shown as grey bands.
The mass splittings of the two $\BMs$ of the linear seesaw are too small to be affected by decoherence, as seen at the lowest two bands.
However, the three $\BMs$ for the inverse seesaw, forming the uppermost three bands, deviate from the grey bands beneath them.
Hence, for a fixed prediction of $R_{ll}$, the value of the squared active-sterile mixing parameters between the genuine and the naive results can vary by up to four orders of magnitude.
The grey area indicates displaced searches by \CMS, and \ATLAS  \cite{ATLAS:2020xyo, CMS:2022fut}.
The shaded grey areas indicate searches for \lnv signals by \CMS and \ATLAS \cite{ATLAS:2015gtp, CMS:2018iaf}.
The two black lines at low masses indicate the reach of the \HLLHC and the \FCC-$ee$ \cite{Drewes:2019fou,Chrzaszcz:2020emg,Blondel:2022qqo}.
The two black lines at high masses indicate the reach of the \LHeC and \FCC-$eh$ \cite{Antusch:2019eiz}.
} \label{fig:Rll of m and theta sqr}
\end{figure}

\resetacronym{BM}%

For the part of the parameter space where heavy neutrinos decay promptly, direct discovery of oscillations by resolving them as proposed in \cite{Antusch:2022hhh} is not possible.
However, the integrated $R_{ll}$ ratio introduced in the last section can still be measured.
Hence, it is relevant to predict this ratio for the realistic \BMs of the linear and inverse seesaw, introduced and motivated in \cite{Antusch:2022ceb} and summarised in \cref{tab:BMs}.
While the heavy neutrino mass splitting in the inverse seesaw scenarios depends on the active-sterile mixing parameter, defined in \eqref{eq:mixing parameter},
\begin{equation}
\Delta m = m_\nu \abs{\vec \theta}^{-2} \,,
\end{equation}
the mass splitting in the linear seesaw scenarios is fixed for each \BM point.
Therefore, in the inverse seesaw, it is always possible to restore coherence and the naive value of $R_{ll}$ by considering larger values of the active-sterile mixing parameter and, accordingly, smaller mass splittings.
However, in the linear seesaw, the damping solely depends on the mass of the heavy neutrinos.
When it becomes relevant, coherence is lost independently of the value of the active-sterile mixing parameter.

The effects of decoherence onto the $R_{ll}$ bands of the linear and inverse seesaw \BM scenarios are depicted in \cref{fig:Rll of m and theta sqr}.
For our baseline estimates of external particle widths, the damping becomes relevant for mass splittings $\Delta m \gtrapprox \unit[1]{eV}$ as shown in \cref{fig:damping numeric}.
The presented \BM points for the inverse seesaw feature such mass splittings for values of the active-sterile coupling in the range of $10^{-4} < \abs{\vec \theta}^2 < 10^{-1}$.
Therefore, the $R_{ll}$ bands deviate from the naive ones in that region.
As discussed above, the bands then follow the contour lines of the damping parameter, such that large damping results in an $R_{ll}$ of one.
The mass splittings for the \BM points for the linear seesaw are such that decoherence effects can be neglected for our baseline estimate of external particle widths, defined in \cref{tab:widths}.
Consequently, the $R_{ll}$ bands do not deviate from the naive ones in this case.

\begin{figure}
\begin{panels}{2}
\includetikz{Rlllowmtheta}
\caption{Scaling of $\sigma_p$ by $\div 100$.} \label{fig:Rll down scaling}
\panel
\includetikz{Rllhighmtheta}
\caption{Scaling of $\sigma_p$ by $\times 100$.} \label{fig:Rll up scaling}
\end{panels}
\caption[Decoherence effect dependence on the external wave packet widths]{
Impact of scaling the proton width $\sigma_p$ on the $R_{ll}$ ratio.
The naive $R_{ll}$ bands are depicted as grey lines, and the coloured lines represent the $R_{ll}$ taking decoherence into account.
Dividing the proton width by a factor of $100$ reduces damping, as shown in panel \subref{fig:Rll down scaling}.
Therefore, the effects in the inverse seesaw are only significant for smaller values of the active-sterile mixing parameter.
Contrary, multiplying the proton width by a factor of $100$ results in enhanced damping, see panel \subref{fig:Rll up scaling}.
The effects in the inverse seesaw are then already significant for larger values of the active-sterile mixing parameter.
With enhanced damping, decoherence becomes relevant also for the linear seesaw with normal ordered light neutrino masses.
Since the mass splitting of the heavy neutrinos in the linear seesaw is fixed, the damping parameter only varies as a function of the mass.
Therefore, damping is relevant for masses in the range $\unit[10]{GeV} \lessapprox m \lessapprox \unit[120]{GeV}$, where the contours in \cref{fig:damping numeric} show a local minimum.
} \label{fig:Rll scaling}
\end{figure}

However, the situation changes if different widths of the external wave packets are assumed.
As shown in \cref{fig:damping dependence}, the most significant effect on the damping is given by varying the proton width $\sigma_p$.
Hence the effect of this variation on the $R_{ll}$ bands are depicted in \cref{fig:Rll scaling}.
If the proton width is divided by $100$, the damping becomes relevant only above mass splittings $\Delta m \gtrapprox \unit[120]{eV}$.
This results in the deviation from the naive $R_{ll}$ only at smaller values of the active-sterile mixing parameter in the inverse seesaw, while the linear seesaw is still not affected as depicted in \cref{fig:Rll down scaling}.
On the contrary, if the proton width is multiplied by $100$, the damping already becomes relevant for mass splittings $\Delta m \gtrapprox \unit[10]{meV}$.
The $R_{ll}$ bands of the inverse seesaw \BM points now deviate from the naive results already for larger values of the active-sterile mixing parameter, and the effects on the linear seesaw \BM points are also relevant, as shown in \cref{fig:Rll up scaling}.
The mass splitting of the linear seesaw with inverted ordering of the light neutrino masses is still too small for decoherence to become relevant.
However, for the normal ordered linear seesaw, damping must be considered.
From \cref{fig:damping dependence}, it can be seen that the mass splitting of the normal ordered linear seesaw results in a small damping for masses $m \lessapprox \unit[10]{GeV}$ and $\unit[120]{GeV} \lessapprox m$.
However, for masses in the range $\unit[10]{GeV} \lessapprox m \lessapprox \unit[120]{GeV}$, corresponding to the local minimum observable in \cref{fig:damping numeric}, decoherence becomes relevant.
The largest damping is around $m \approx \unit[50]{GeV}$, and thus the observed $R_{ll}$ deviates from the naive one in this region.
Since, as discussed above, coherence cannot be restored by varying the active-sterile mixing parameter as in the case of the inverse seesaw, the observed $R_{ll}$ is close to one in this mass range for all values of the active-sterile mixing angle.
Therefore, the $R_{ll}$ bands form a pole around $m \approx \unit[50]{GeV}$ as shown in \cref{fig:Rll up scaling}.

The effects of the damping parameter are crucial when reinterpreting prompt searches for \lnv signals.
Regions that result in an $R_{ll}$ close to zero for a considered model do not yield any \lnv events and, thus, are not restricted by these searches.
However, if damping is significant, the true value of $R_{ll}$ might significantly deviate from the naive one, as discussed above.
Therefore, decoherence effects can result in prompt searches for \LNV becoming applicable in regions of the parameter space that seem unconstrained when neglecting decoherence.

\section{Conclusion} \label{sec:conclusion}

\resetacronym{NNO}

Low-scale symmetry-protected seesaw models generically predict the appearance of pseudo-Dirac pairs of heavy neutrinos.
It is usually expected that for such models, \LNV is severely suppressed \cite{Kersten:2007vk}.
However, these considerations omit the crucial possibility of \NNOs which can lead to a sizable number of \lnv events if the oscillation period is shorter or of the same order as the lifetime of the heavy neutrinos \cite{Antusch:2022ceb}.
In particular, for long-lived heavy neutrinos, the oscillation pattern between \lnc and \lnv events may be reconstructible in collider experiments \cite{Antusch:2022hhh}.
Apart from \NNOs, decoherence can yield observable amounts of \LNV by reducing the destructive interference between propagating mass eigenstates and, with it, the suppression of \LNV.
While this obstructs the reconstruction of the oscillation pattern, it also enhances the chances of observing \LNV when the heavy neutrino's lifetime is smaller than its oscillation period.
In this work, we have quantitatively studied this effect for the first time.

To this end, we have derived oscillation probabilities for \NNOs in the framework of \QFT with external wave packets, extending previous results, see \cite{Beuthe:2001rc}, with a reformulation that depends on the time difference between production and decay of the heavy neutrinos.
The derivations presented here are not only technically simpler than results that depend on the distance between production and decay of the heavy neutrinos but also more readily applicable to cases where an interplay between the oscillation period and lifetime is relevant, such as the \lnv over \lnc event ratio $R_{ll}$ and the reconstruction of the \NNO pattern, which requires a translation of the oscillations into the proper time frame of the heavy neutrinos \cite{Antusch:2017ebe,Antusch:2020pnn,Antusch:2022hhh}.
The analytical calculation relies on expansions in small parameters and differentiates between the \NDR and the \TDR that apply for short and longer-lived particles.
The numerical comparison shows that the time-dependent results for the \NDR and \TDR are almost identical, such that a smooth transition connects the two regimes.
On the contrary, in the distance-dependent results, the differences between the two regimes are significant such that no smooth transition is possible.
We have compared the time-dependent analytical results with a numeric calculation using \MC data and confirmed a broad range of applicability.

In \cite{Antusch:2022ceb}, we proposed using a single damping parameter $\lambda$ that encodes all decoherence effects in \NNOs.
Here, we provide the formulae necessary to calculate this damping parameter from first principles as a function of the heavy neutrinos' mean mass and mass splitting.
To that end, we provide the conditions under which the dependence on the decay width $\Gamma$ and, therefore, the dependence on the propagation time between the production and detection of the heavy neutrinos can be neglected.
Based on our analytical results, we conjecture that the dependence of $\lambda$ on $m$ and $\Delta m$ also extends to $\Gamma$ larger than the ones we calculated numerically.

Furthermore, we discussed a novel time-independent contribution to the phase and derived analytical formulae for this phase shift in the \NDRlong and \TDRlong.
However, in the parameter region under consideration, the phase shift can be neglected since it is either small or damping is large, which results in a suppression of the phase shift effects.

The employed framework depends on the widths of the external particles' wave packets in position space.
These input parameters need to be adjusted to the described experimental situation.
We have picked well-motivated values in order to present our results.
Additionally, we have discussed the dependence of relevant quantities on changes in these parameters.
We have also discussed possible limitations of applicability of the formalism from the \JS time threshold.
In our computations, we followed a conservative approach restricted to $\Gamma < \Gamma_{\JS} = \unit[0.1]{eV}$.

We illustrate the impact of decoherence by presenting bands of $R_{ll}$ in the ($m$, $\abs{\vec \theta}^2$)-parameter plane and show significant deviations from the predictions when decoherence is neglected.
Hence, we quantify for the first time how the transition of a coherently oscillating pseudo-Dirac pair to two independently acting Majorana particles affects the phenomenology and, therefore, the discovery prospects of symmetry-protected low-scale seesaw models.

From the results of this work, it is clear that the possibility of decoherence has to be considered when studying \lnv signatures.
Large decoherence not only suppresses the oscillation pattern but also disables the mechanism that suppresses \LNV for a pseudo-Dirac pair.
Therefore, the phenomenology of a pseudo-Dirac pair changes significantly in regions of parameter space that exhibit sizable decoherence.

\subsection*{Acknowledgements}

The work of Jan Hajer was partially supported by the Portuguese Fundação para a Ciência e a Tecnologia (FCT) through the projects CFTP-FCT unit UIDB/\allowbreak00777/\allowbreak2020 and UIDP/\allowbreak00777/\allowbreak2020.

\appendix

\section{Momentum integration} \label{sec:momentum integration}

\resetacronym{STE}
\resetacronym{NDR}
\resetacronym{TDR}
\resetacronym{LDR}

The integration of the transition amplitude \eqref{eq:time integrated amplitude} over the three-momentum of the intermediate particle is performed using different techniques depending on how fast the complex phase varies over the width of the intermediate particles' wave packet.
In the \NDR, the phase varies slowly such that the integral is approximated around the maximum of the intermediate wave packet up to the second order in the momentum.
This method is referred to as Laplace's method.
If the phase varies rapidly over the intermediate wave packet, the method of stationary phase is used, where the largest contribution to the integral is obtained near the point for which the phase has an extremum.
In the \TDR, the phase varies slowly in directions transversal to the reconstructed momentum $\vec p_0$, while in the longitudinal direction, Laplace's method can still be used.
The third regime, called the \LDR, in which the method of stationary phase has to be used for all directions of the momentum $\vec p$, is not relevant to the phenomenology considered in this work.
In the following, we treat the \NDR and the \TDR separately and derive an oscillation probability in each of them.

\subsection{\sentence\NDRlong} \label{sec:ND momentum integration}

For short times the phase \eqref{eq:time integrated phase and decay} varies slowly, as a function of the momentum, over the size of the intermediate particles' wave packet given by the \EME \eqref{eq:time integrated EME}.
Thus the momentum integration can be performed using Laplace's method, where the integral is approximated around the momentum $\vec p_i$ for which the \EME together with the decay term \eqref{eq:time integrated phase and decay} is minimal.
Hence a necessary condition that needs to be fulfilled in order to apply Laplace's method is
\begin{equation}
\eval*{\dv{\vec p}\left(f_i(\vec p) + \gamma_i(t,\vec p)\right)}_{\vec p = \vec p_i} = 0 \,.
\end{equation}
While the \EME yields a Gauss-like shape for the amplitude with a maximum near the reconstructed momentum, the decay term favours large momenta.
Therefore we are interested in the minima of the \EME.
For the analytic computation, it is assumed that the impact of the decay term on the position of the maximum is negligible, which can be justified if the decay term varies slowly over the width of the intermediate particles' wave packet.
The condition for this assumption to be valid is derived in \eqref{eq:NDR decay condition 2}.
However, in the numerical computation in \cref{sec:pseudocode}, the effects of the decay term are taken into account.
With the decay term neglected, the position of the maximum is expanded around the maximum of the reconstructed momentum
\begin{equation} \label{eq:NDR mass splitting expansion}
\vec p_i = \vec p_0 + \vec p_1 \delta_i + \order*{\delta_i^2} \,.
\end{equation}
where the mass splitting expansion parameter and the \NLO term are
\begin{align} \label{eq:NDR mass splitting expansion parameter}
\delta_i &= \frac{m_i^2 - m_0^2}{2 E_0^2} \,, &
\vec p_1 &= E_0 \mat \Sigma_0^{-1} \left(\frac{\vec u_P^{}}{2 \sigma_{EP}^2} + \frac{\vec u_D^{}}{2 \sigma_{ED}^2}\right) \,, &
\vec u_V &= \vec v_V^{} - \vec v_0 \,, &
V &\in \{P,D\} \,,
\end{align}
The matrix $\mat \Sigma_0^{-1}$ appearing in the \NLO term of the expansion is the inverse of the Hessian of the \EME at \LO in the momentum expansion, which is given in \eqref{eq:NDR EME Hessian}.
Its eigenvalues can be interpreted as effective widths of the wave packet of the intermediate particle mass eigenstates in different directions, see \cref{sec:dispersion regimes}.
The corresponding expansion of the energy of the mass eigenstate at the minimum of the \EME around the reconstructed energy yields
\begin{align} \label{eq:NDR energy-mass splitting expansion}
E_i &= E_0 + E_1 \delta_i + \order*{\delta_i^2} \,, &
E_1 &= \vec p_1 \dot \vec v_0 + E_0 \,, &
E_i &= E_i(\vec p_i) \,.
\end{align}
Finally, the expansion of the velocity of the mass eigenstate at the minimum reads
\begin{align} \label{eq:NDR velocity-mass splitting expansion}
\vec v_i &= \vec v_0 + \vec v_1 \delta_i + \order*{\delta_i^2} \,, &
\vec v_1 &= \frac{\vec p_1 - \vec v_0 E_1}{E_0} \,, &
\vec v_i &= \vec v_i(\vec p_i) \,.
\end{align}

\subsubsection{Expansion} \label{sec:ND expansion}

In order to evaluate the momentum integral in the amplitude \eqref{eq:time integrated amplitude}, each exponential term is expanded up to second order in the momentum $\vec p$ around the minimum of the \EME at $\vec p_i$ resulting in a function that can be evaluated using Gaussian integration as demonstrated in \cref{sec:ND momentum integral}.

\paragraph{\sentence\EMElong}

The expansion of the \EME \eqref{eq:time integrated EME} around its minimum at $\vec p_i$ results in
\begin{equation} \label{eq:NDR EME expansion}
f_i(\vec p) = f_i + \frac12 (\vec p - \vec p_i)^\trans \mat \Sigma_i (\vec p - \vec p_i) + \order*{\abs{\vec p - \vec p_i}^3} \,,
\end{equation}
where the linear term vanishes since the evaluation is performed at the minimum.
The constant term and the Hessian of the \EME at the minimum read
\begin{align} \label{eq:NDR EME parameters}
f_i &= f_i(\vec p_i) \,, &
\mat \Sigma_i &= \mat \Sigma_0 + \order*{\delta_i} \,,
\end{align}
where the constant term to \LO in the mass splitting expansion defines the mass width
\begin{align}
f_i &= f_1 \delta_i^2 + \order*{\delta_i^4} \,, &
f_1 &= \abs*{\frac{\vec p_1}{2 \sigma_{\vec pP}^{}}}^2 + \left[\frac{e^{}_{1P}}{2 \sigma_{EP}^{}}\right]^2 + (P\to D) \,, &
\sigma_m &:= \frac{E_0}{2 \sqrt{f_1}}
\end{align}
where
\begin{align}
e^{}_{iV} &:= e^{}_{iV}(\vec p_i) = e^{}_{1V} \delta_i \,, &
e^{}_{1V} &:= E_1 - \vec p_1 \dot \vec v_V \,.
\end{align}
and the \LO term in the mass splitting expansion of the Hessian is
\begin{equation}\label{eq:NDR EME Hessian}
\mat \Sigma_0 = \frac{\identity}{2 \sigma_{\vec pP}^2} + \frac{\vec u_P^{} \otimes \vec u_P^{}}{2 \sigma_{EP}^2} + \left(P\to D\right) \,.
\end{equation}
From the inequality \eqref{eq:width estimate} follows that the Hessian can be estimated to take values within
\begin{equation}
\frac{1}{2\sigma_{\vec p}^2} \lessapprox \abs{\mat \Sigma_i} \lessapprox \frac{1}{2\sigma_E^2} \,,
\end{equation}
where $\abs{\mat \Sigma_i}$ denotes that the considered inequality has to hold for all eigenvalues of $\mat \Sigma_i$.

\subparagraph{\JSlong integration}

Solving the energy integral via the \JS theorem as demonstrated in \cref{sec:JS} introduces the complex pole energy \eqref{eq:pole energy}.
The dependence on the mass eigenstate energy \eqref{eq:pole energy dependence} can then be used to estimate the effects of the decay width expansion parameter \eqref{eq:decay width expansion parameter} on the \EME
\begin{equation} \label{eq:EME energy correction}
f_i(E_i^\prime(\vec p),\vec p) = \left[\frac{e^{}_P(E_i^\prime(\vec p), \vec p)}{2 \sigma^{}_{EP}}\right]^2 + \dots \,,
\end{equation}
where
\begin{equation}
e^{}_P(E_i^\prime(\vec p), \vec p) = \sqrt{1 - 2 \i \epsilon_i(\vec p)} \sqrt{\abs{\vec p}^2 + m_i^2} - E_0 - (\vec p - \vec p_0) \dot \vec v_P \,.
\end{equation}
and the ellipses denote terms that do not depend on the decay width expansion parameter, and for simplicity, we only consider the term at production, keeping in mind that the same arguments hold for the equivalent term at detection.
Furthermore, assuming that the energy and mass splitting expansion parameters are of the same order, the constant term and the Hessian of the momentum expanded \EME \eqref{eq:NDR EME parameters} are to \LO
\begin{align}
f_i &= \left(\frac{\delta_i - \i \epsilon_i(\vec p)}{2 \sigma_{EP}^{}} E_0\right)^2 + \dots \,, &
\mat \Sigma_i &= \left(\frac{\vec u_P^{}}{2 \sigma_{EP}^{}}\right)^2 + \order*{\delta_i} + \order*{\epsilon_i^2(\vec p)} + \dots \,.
\end{align}
The effects of the decay width expansion parameter on the Hessian are subleading and can therefore be neglected.
While the effects on the \EME are relevant, the term itself does only contribute to the damping of the amplitude \eqref{eq:time integrated amplitude} if it differs greatly between different mass eigenstates.
This is due to the normalisation discussed in \cref{sec:NNO probability}.
The effects of the decay width expansion parameter are therefore neglected in the analytical derivation by using \eqref{eq:time integrated EME} while they are taken into account in the numerical calculations in \cref{sec:pseudocode}.

\paragraph{Phase}

The expansion of the phase \eqref{eq:time integrated phase and decay} around the minimum of the \EME results in
\begin{equation} \label{eq:NDR phase expansion}
\phi_i(t,\vec x,\vec p) = \phi_i(t,\vec x) + \vec \Delta_i(t, \vec x) \dot (\vec p - \vec p_i) + \frac12 (\vec p - \vec p_i)^\trans \mat R_i(t) \, (\vec p - \vec p_i) + \order*{\abs{\vec p - \vec p_i}^3} \,,
\end{equation}
where the constant term, the linear coefficient, the Hessian, and the velocity of the $i$-th mass eigenstate are
\begin{align} \label{eq:NDR phase parameters}
\phi_i(t,\vec x) &= \phi_i(t,\vec x,\vec p_i) \,, &
\vec \Delta_i(t, \vec x) &= \vec v_i t - \vec x \,, &
\mat R_i(t) &= \frac{\identity - \vec v_i \otimes \vec v_i}{E_i} t \,, &
\vec v_i &= \frac{\vec p_i}{E_i} \,.
\end{align}
Since the inverse of $\mat \Sigma_i$ can be interpreted as an effective width of the intermediate wave packet \eqref{eq:NDR inverse width,eq:TDR inverse width}, it can be used to quantify the condition that the phase varies slowly over the width of the intermediate wave packet.
Replacing $\abs{\vec p - \vec p_i}^2$ in the expansion series of the phase \eqref{eq:NDR phase expansion} with $2 \abs{\mat \Sigma_i}^{-1}$ and requiring that the linear and quadratic terms are small results in
\begin{align} \label{eq:NDR phase condition}
2 \abs{\vec \Delta_i(t, \vec x)}^2 &\ll \abs{\mat \Sigma_i} \,, &
\abs{\mat R_i(t)} &\ll \abs{\mat \Sigma_i} \,,
\end{align}
These conditions can be approximated to read
\begin{align} \label{eq:NDR phase condition 2}
2\abs{\vec v_i t - \vec x}^2 &\ll \abs{\mat \Sigma_0} \,, &
\frac{\abs{\identity - \vec v_i \otimes \vec v_i}}{E_i} t &\ll \abs{\mat \Sigma_0} \,.
\end{align}
Since the averaging over the distance (or time) in a later step ensures that $\vec v_i t \approx \vec x$, the linear condition is automatically satisfied.
The quadratic condition can be approximated as
\begin{equation}
\frac{t}{E_i} \ll \frac{1}{2 \sigma_{\vec p}^2} \,,
\end{equation}
where the eigenvalue of $\mat \Sigma_0$ containing $\sigma_{\vec p}^2$ is used since it imposes the most restrictive condition, see \eqref{eq:smallest eigenvalue}.
For the same reason, the velocity-dependent parts in the contribution from the phase can be neglected.
This condition defines the short-time threshold up to which this integration method is valid.
It is given by
\begin{align} \label{eq:NDR short-time threshold}
t_i^\text{short} &= t^\text{short} + \order*{\delta_i} = E_i \abs{\mat \Sigma_i}_\text{smallest} \,, &
t^\text{short} &= \frac{E_0}{2 \sigma_{\vec p}^2} \,, &
t &\lessapprox t^\text{short} \,,
\end{align}
and it is usually sufficient to work with the \LO approximation.

\paragraph{Decay term}

The decay term \eqref{eq:time integrated phase and decay} is also expanded up to second order in its momenta around the minimum of the \EME yielding
\begin{equation} \label{eq:NDR decay term expansion}
\gamma_i(t,\vec p) = \gamma_i(t) - \vec \digamma_i \dot (\vec p - \vec p_i) + \frac12 (\vec p - \vec p_i)^\trans \mat W_i(t) \, (\vec p - \vec p_i) + \order*{\abs{\vec p - \vec p_i}^3} \,,
\end{equation}
where the constant term, the linear coefficient, the Hessian and the parameter appearing in \eqref{eq:decay width expansion parameter} evaluated at the minimum of the \EME are given by
\begin{align} \label{eq:NDR decay term parameters}
\gamma_i(t) &= \gamma_i t \,, &
\vec \digamma_i(t) &= \vec v_i \epsilon_i t \,, &
\mat W_i(t) &= \frac{3 \vec v_i \otimes \vec v_i - \identity}{E_i} \epsilon_i t \,, &
\gamma_i &= \frac{m_i \Gamma_i}{2 E_i} \,, &
\epsilon_i &= \frac{\gamma_i}{E_i} \,.
\end{align}
Similar to the conditions appearing in the evaluation of the phase \eqref{eq:NDR phase condition}, two conditions can be derived, ensuring that the decay term varies slowly over the width of the wave packet
\begin{align} \label{eq:NDR decay condition}
2 \abs{\vec \digamma_i(t)}^2 &\ll \abs{\mat \Sigma_i} \,, &
\abs{\mat W_i(t)} &\ll \abs{\mat \Sigma_i} \,.
\end{align}
For times earlier than the short-time threshold \eqref{eq:NDR short-time threshold}, these conditions become
\begin{align} \label{eq:NDR decay condition 2}
\abs{\vec p_i} \epsilon_i &\ll \sigma_{\vec p} \,, &
\epsilon_i &\ll \frac{1}{\abs*{3 \vec v_i \otimes \vec v_i - \identity}} \,.
\end{align}
The linear condition holds as long as the decay width is of the same order or smaller as the momentum width, while the quadratic condition is satisfied for the assumptions used in the decay width expansion during the \JS integration in \cref{sec:JS}.

\subsubsection{Integration} \label{sec:ND momentum integral}

The momentum integral can then be evaluated using a standard technique for multidimensional Gaussian integrals over the coordinates $\vec x$ with a symmetric positive definite matrix $\mat A$ and a linear term $\vec b$
\begin{equation} \label{eq:Gauss integral}
\int[\d^3 \vec x] \exp\left[\vec b \dot \vec x - \frac12 \vec x^\trans \mat A \vec x\right] = \exp\left[\frac12\vec b^\trans \mat A^{-1} \vec b\right] \,.
\end{equation}
The vector $\vec b$ contains the linear order coefficients appearing in the momentum expansion of the phase \eqref{eq:NDR phase expansion} and the decay term \eqref{eq:NDR decay term parameters} and reads
\begin{equation}
\vec b = \i \vec \Delta_i(t, \vec x) + \vec \digamma_i(t) = \i \vec \Delta_i(t, \vec x) + \order*{\epsilon_i} \,,
\end{equation}
For the analytical derivation, we neglect the correction from the decay term and proceed solely with the linear coefficient from the expansion of the phase around the minimum of the energy-momentum integral.
In the numerical calculation presented in \cref{sec:pseudocode} the minimum $\vec p_i$ is computed for the sum of the \EME and the decay term, such that the linear contribution $\vec \digamma_i(t)$ is absent all together.
The matrix $\mat A$ collects the Hessian matrices resulting from the momentum expansion of the \EME \eqref{eq:NDR EME parameters}, the phase \eqref{eq:NDR phase parameters}, and the decay term \eqref{eq:NDR decay term parameters} around the minimum of the \EME and reads
\begin{equation} \label{eq:NDR matrix}
\mat A = - \mat \Sigma_i - \mat W_i(t) - \i \mat R_i(t) = - \mat \Sigma_i + \order*{\frac{t}{t^\text{short}}} \,.
\end{equation}
For times earlier than the short-time threshold \eqref{eq:NDR short-time threshold}, it is justified to approximate this sum with just the contribution from the \EME, see the conditions \eqref{eq:NDR phase condition 2,eq:NDR decay condition}.

Hence the Hessian from the \EME expansion \eqref{eq:NDR EME parameters} together with the linear term of the phase expansion \eqref{eq:NDR phase expansion} integrated over the momentum using the Gaussian integral \eqref{eq:Gauss integral} yield the \STE
\begin{equation} \label{eq:NDR STE}
F_i(t,\vec x) := \frac12 \vec \Delta_i^\trans(t, \vec x) \mat \Sigma_i^{-1} \vec \Delta_i(t, \vec x) \,.
\end{equation}
For a given time and velocity, this term leads to an exponential damping of the transition amplitude for values of $\vec x$ \emph{far} from $\vec v_i t$.
Here, \emph{far} is defined by the eigenvalues of $\mat \Sigma_i$.
We have explicitly discussed the effects of the decay width expansion parameter after \eqref{eq:EME energy correction} and found the \STE to agree with \cite{Beuthe:2001rc}.

The oscillation amplitude in the \NDR after momentum integration is therefore given by
\begin{equation} \label{eq:NDR transition amplitude}
\mathcal A_i(t,\vec x) \propto \exp\left[- f_i - \gamma_i(t) - F_i(t,\vec x) - \i \phi_i(t,\vec x)\right] \,.
\end{equation}
where the terms in the exponential are the constant \EME \eqref{eq:NDR EME parameters}, the time dependent decay term \eqref{eq:NDR decay term parameters}, the spacetime dependent \STE \eqref{eq:NDR STE}, and the spacetime dependent phase \eqref{eq:NDR phase parameters}.

\subsection{\sentence\TDRlong} \label{sec:TD momentum integration}

For times later than the short-time threshold \eqref{eq:NDR short-time threshold}, derived in the previous section, the oscillations are fast compared to the width of the wave packet such that Laplace's method is no longer the preferred method for evaluating the momentum integral.
However, this argument depends on the direction of the heavy neutrino momentum.
The second order term of the phase \eqref{eq:NDR phase expansion} yields a contribution $\flatfrac{\abs{\vec p}^2 t}{E}$ for momenta orthogonal to $\vec v_i$, while momenta in the direction of $\vec v_i$ obtain an additional Lorentz contraction factor that leads to $\flatfrac{(1 - \vec v_i^2) \abs{\vec p}^2 t}{E}$.
Since at \LO $\vec v_i = \vec v_0$, this factor slows down the oscillations in the direction of the reconstructed momentum.
Laplace's method is therefore still preferred in the direction along $\vec p_0$, while in directions orthogonal to it, the method of stationary phase is used.

\subsubsection{Stationary phase}

The linear term of the expanded phase \eqref{eq:NDR phase expansion} averages oscillations to zero for momenta transversal to $\vec x$.
Therefore it can be assumed that $\vec p$ is parallel to $\vec x$.
Additionally, the \EME requires that $\vec p$ is parallel to $\vec p_0$, which yields that $\vec x$ has to be parallel to $\vec p_0$.
Assuming that $\vec p_0$ is in the $z$ direction the method of stationary phase can be used for $\eval{\vec x}_x$ and $\eval{\vec x}_y$ which yields
\begin{equation}
\pdv{p_x} \phi(t,\vec x,\vec p) = \pdv{p_y} \phi(t,\vec x,\vec p) = 0 \,,
\end{equation}
resulting in $p_x = p_y = 0$.
Using those, the argument of the exponential of the amplitude \eqref{eq:initial amplitude} does not depend on $p_x$ and $p_y$ anymore such that only the integration over $\proj p = \eval{\vec p}_z$ is left.
This integration is done using Laplace's method, as in \cref{sec:ND momentum integration}.

The argument of the \EME \eqref{eq:EME} can then be expressed as
\begin{equation} \label{eq:TDR EME}
f_i(E_i(\proj p), \proj p) = \left(\frac{\proj p - \proj p_0}{2 \sigma_{\vec pP}^{}}\right)^2 + \left(\frac{e^{}_{iP}(\proj p)}{2 \sigma_{EP}^{}}\right)^2 + (P\to D) \,,
\end{equation}
where
\begin{equation}
e^{}_{iV}(\proj p) = E_i(\proj p) - E_0 - (\proj p - \proj p_0) \proj v_V \,.
\end{equation}
Similar to the \NDR, the phase, the \EME, and the decay term are all expanded around the momentum $\proj p_i$ for which the \EME is minimal.
The effects of the decay term onto the position of the maximum are neglected in the analytical derivation, and the exact conditions for this approximation to hold are given in \eqref{eq:TDR decay condition 2}.
The position of the maximum up to linear order in the mass splitting expansion parameter \eqref{eq:NDR mass splitting expansion parameter} is
\begin{align} \label{eq:TDR mass splitting expansion}
\proj p_i &= \proj p_0 + \proj p_1 \delta_i + \order*{\delta_i^2} \,, &
\proj p_1 &= \frac{E_0}{\proj \Sigma_0} \left(\frac{\proj u_P}{2\sigma_{EP}^2} + \frac{\proj u_D}{2\sigma_{ED}^2}\right) \,, &
\proj u_V &:= \proj v_V^{} - \proj v_0 \,,
\end{align}
the mass splitting expansion of the mass eigenstate energy reads
\begin{align} \label{eq:TDR energy-mass splitting expansion}
E_i &= E_0 + E_1 \delta_i + \order*{\delta_i^2} \,, &
E_1 &= E_0 + \proj p_1 \proj v_0 \,, &
E_i &= E_i(\proj p_i)
\,,
\end{align}
and the mass splitting expansion of the mass eigenstate velocity is
\begin{align} \label{eq:TDR velocity-mass splitting expansion}
\proj v_i &= \proj v_0 + \proj v_1 \delta_i + \order*{\delta_i^2} \,, &
\proj v_1 &= \frac{\proj p_1 - E_1 \proj v_0}{E_0} \,, &
\proj v_i &= \proj v_i(\proj p_i) \,.
\end{align}
The effects of the decay width expansion parameter are neglected here for the same reasons as in \cref{sec:ND expansion}.

\subsubsection{Expansion}

Just as in the \NDR, the terms in the exponent of the amplitude \eqref{eq:time integrated amplitude} need to be expanded up to second order in the momentum in order to perform a Gaussian integration in \cref{sec:TD integration}.

\paragraph{\sentence\EMElong}

The expansion of the \EME \eqref{eq:TDR EME} around the momentum of the minimum is given by
\begin{equation} \label{eq:TDR EME expansion}
f_i(E_i(\proj p), \proj p) = f_i + \frac12 \proj \Sigma_i \left(\proj p - \proj p_i\right)^2 + \order[\big]{\abs*{\proj p - \proj p_i}^3} \,,
\end{equation}
where the linear term vanishes since the expansion is evaluated at the minimum, while the constant term and the Hessian are given by
\begin{align} \label{eq:TDR EME parameters}
f_i &= f_1 \delta_i^2 + \order*{\delta_i^4} = f_i(E_i, \proj p_i)\,, &
\proj \Sigma_i &= \proj \Sigma_0 + \order*{\delta_i} \,,
\end{align}
and their \LO contributions in the mass splitting expansion are given by
\begin{align} \label{eq:TDR EME Hessian}
f_1 &= \left(\frac{\proj p_1}{2 \sigma_{\proj pP}^{}}\right)^2 + \left(\frac{E_1 - \proj p_1 \proj v_P}{2 \sigma_{EP}^{}}\right)^2 + (P\to D) \,, &
\proj \Sigma_0 &= \frac{1}{2\sigma_{\vec pP}^2} + \frac{\proj u_P^2}{2\sigma_{EP}^2} + (P\to D) \,.
\end{align}

\paragraph{Phase}

The expansion of the phase \eqref{eq:time integrated phase and decay} around the momentum of the minimum of the \EME yields
\begin{equation} \label{eq:TDR phase expansion}
\phi_i(t, \proj x, \proj p) = \phi_i(t, \proj x) + \proj \Delta_i(t, \proj x) \left(\proj p - \proj p_i\right) + \frac12 \proj R_i(t) (\proj p - \proj p_i)^2 + \order[\big]{\abs*{\proj p - \proj p_i}^3} \,,
\end{equation}
where the constant term, the linear coefficient, the Hessian, and the velocity of the mass eigenstate are given by
\begin{align} \label{eq:TDR phase parameters}
\phi_i(t, \proj x) &= \phi_i(t, \proj x, \proj p_i) \,, &
\proj \Delta_i(t, \proj x) &= \proj v_i t - \proj x \,, &
\proj R_i(t) &= \frac{m_i^2}{E_i^3} t \,, &
\proj v_i &= \frac{\proj p_i}{E_i} \,.
\end{align}
Similar to the case of the \NDR, a time threshold can be obtained by requiring that the phase varies slowly over the width of the wave packet.
The wave packet width is approximated by $\proj \Sigma_i^{-\nicefrac12}$ and used to reexpress the momentum deviations.
The conditions resulting from the requirement that the linear and quadratic terms are small are
\begin{align} \label{eq:TDR phase condition}
2 \proj \Delta_i^2(t, \proj x) &\ll \proj \Sigma_i \,, &
\proj R_i(t) & \ll \proj \Sigma_i \,.
\end{align}
The linear condition is ensured by the distance average performed in the next section, while the quadratic condition defines the long-time threshold.
A time threshold independent of the mass eigenstates can be defined by approximating it at \LO in the mass splitting expansion \eqref{eq:NDR mass splitting expansion}, leading to the validity range of the \TDR
\begin{align} \label{eq:TDR long-time threshold}
t_i^\text{long} &= t^\text{long} + \order*{\delta_i} =
\proj \Sigma_i \frac{E_i^3}{m_i^2} \,, &
t^\text{long} &:= \proj \Sigma_0 \frac{E_0^3}{m_0^2} \,, &
t^\text{short} &\lessapprox t \lessapprox t^\text{long} \,.
\end{align}

\paragraph{Decay term}

The expansion of the decay term \eqref{eq:time integrated phase and decay} is the same as in the \NDR \eqref{eq:NDR decay term expansion} where all vector quantities are replaced by their corresponding longitudinal component.
\begin{equation} \label{eq:TDR decay term expansion}
\gamma_i(t,\proj p) = \gamma_i(t) - \proj \digamma_i \dot (\proj p - \proj p_i) + \frac12 (\proj p - \proj p_i)^\trans \proj W_i(t) \, (\proj p - \proj p_i) + \order*{\abs{\proj p - \proj p_i}^3} \,.
\end{equation}
The constant term, the linear coefficient, the Hessian and the terms appearing in \eqref{eq:decay width expansion parameter} are given by
\begin{align} \label{eq:TDR decay term parameters}
\gamma_i(t) &= \gamma_i t \,, &
\gamma_i &= \frac{m_i \Gamma_i}{2 E_i} \,, &
\proj \digamma_i(t) &= \proj v_i \epsilon_i t \,, &
\proj W_i(t) &= \frac{3 \proj v_i^2 - 1}{E_i} \epsilon_i t \,, &
\epsilon_i = \frac{\gamma_i}{E_i} \,.
\end{align}
The conditions for the decay term to vary slowly over the width of the neutrino wave packet read
\begin{align} \label{eq:TDR decay condition}
2 \proj \digamma_i^2(t) &\ll \proj \Sigma_i \,, &
\proj W_i(t) &\ll \proj \Sigma_i \,.
\end{align}
assuming $t \lessapprox t_i^\text{long}$, these conditions reduce to
\begin{align} \label{eq:TDR decay condition 2}
\epsilon_i \proj p_i \frac{E_i^2}{m_i^2} &\ll \sigma_E^{} \,, &
\left(3 \proj v_i^2 - 1\right) \epsilon_i &\ll \frac{m_i^2}{E_i^2} \,.
\end{align}
Similar to the situation in the \NDR \eqref{eq:NDR decay condition 2}, the quadratic condition is automatically satisfied.
The linear condition requires the decay width to be of the same order or smaller than the effective momentum width of the wave packet, which is typically of the order of the energy width $\sigma_E^{}$ and thus much smaller compared to the width in the \NDR.
Additionally, the factor $\flatfrac{E_i}{m_i}$ can result in a violation of the condition for ultra-relativistic particles.
This reflects the fact that for such particles, the \TDR is valid up to arbitrarily large times, such that eventually, the decay term becomes important.
In the numerical estimation of the damping, the effect of the decay term is taken into account.

\subsubsection{Integration} \label{sec:TD integration}

The integral in the transition amplitude \eqref{eq:time integrated amplitude} can be solved using the general result \eqref{eq:Gauss integral} where the vector $\vec b$, now scalar, contains the first order terms from the expansion of the phase \eqref{eq:TDR phase expansion} and the decay term \eqref{eq:TDR decay term expansion} and reads
\begin{equation}
b = \i \proj \Delta_i(t, \proj x) + \proj \digamma_i(t) = \i \proj \Delta_i(t, \proj x) + \order*{\epsilon_i} \,.
\end{equation}
In the following, the decay width expansion parameter correction is neglected for the analytical derivation but taken into account for the numerical calculation.
The Hessians resulting from the momentum expansion of the \EME \eqref{eq:TDR EME expansion}, the phase \eqref{eq:TDR phase expansion}, and the decay term \eqref{eq:TDR decay term expansion} are collected in $\mat A$ which is now a scalar and reads
\begin{equation} \label{eq:TDR matrix}
A = - \proj \Sigma_i - \proj W_i(t) - \i \proj R_i(t) = - \proj \Sigma_i + \order*{\frac{t}{t^\text{long}}} \,.
\end{equation}
For times earlier than the long-time threshold \eqref{eq:TDR long-time threshold}, the \LO contribution is given by the Hessian of the \EME, see conditions \eqref{eq:TDR phase condition,eq:TDR decay condition}.
For the numerical results, the contribution from the Hessian of the phase and the decay term are taken into account.

The integration \eqref{eq:Gauss integral} over the momentum yields the \STE
\begin{equation} \label{eq:TDR STE}
F_i(t, \proj x) := \frac{\proj \Delta_i^2(t, \proj x)}{2 \proj \Sigma_i(t)}  \,,
\end{equation}
and the amplitude \eqref{eq:initial amplitude} after integration is given by
\begin{equation} \label{eq:TDR amplitude}
\mathcal A_i(t, \proj x) \propto \exp\left[- f_i - \gamma_i(t) - F_i(t, \proj x) - \i \phi_i(t, \proj x)\right] \,,
\end{equation}
where the terms in the exponent are the \EME \eqref{eq:TDR EME parameters}, the decay term \eqref{eq:TDR decay term parameters}, the \STE\eqref{eq:TDR STE}, and the phase \eqref{eq:TDR phase parameters}.

\section{Distance integration} \label{sec:distance integration}

In order to obtain an oscillation probability, the amplitudes for different mass eigenstates are coherently summed over
\begin{equation} \label{eq:oscillation probability spacetime}
\mathcal P(t,\vec x) \propto \sum_{ij} \mathcal A_i(t,\vec x) \mathcal A_j^*(t,\vec x) \,.
\end{equation}
An average over the oscillation distance is performed to obtain a formula dependent on the time.
For the \NDR as well as for the \TDR the only distance-dependent components in the amplitudes \eqref{eq:NDR transition amplitude,eq:TDR amplitude} are the phases \eqref{eq:NDR phase parameters,eq:TDR phase parameters} and the \STEs \eqref{eq:NDR STE,eq:TDR STE}.
Since the \STE restricts the values of $t$ and $\vec x$ to a range, similar to how the \EME \eqref{eq:EME} restricts the values of $p$ and $E$, the probability is expanded up to second order around $\vec x_{ij}$, the position of the minimum of the \STE, before the resulting Gauss-like integral is evaluated.

Typically, neither the oscillation distance $\vec x$ nor the oscillation time $t$ are measured with perfect precision.
Therefore, the oscillation probability \eqref{eq:oscillation probability spacetime} is either integrated over a span of time, as in \cite{Beuthe:2001rc}, or over a region of space in order to address the experimental uncertainty.
In the following, the second possibility is employed, which yields an oscillation probability as a function of time.
The distance-integrated oscillation probability reads
\begin{equation} \label{eq:oscillation probabillity distance average}
\mathcal P(t) \propto \int[\d \vec x]_{\vec x_0 - \Delta \vec x}^{\vec x_0 + \Delta \vec x} \mathcal P(t,\vec x) \,,
\end{equation}
where $\Delta \vec x$ labels the experimental uncertainty in determining the distance travelled by the intermediate particle between its production and decay.
This distance has to be larger than the width of the intermediate particle in spacetime, which describes the minimal uncertainty due to the wave packet nature of the intermediate particle.
Due to the exponential damping stemming from the \STEs \eqref{eq:NDR STE,eq:TDR STE} for values $\abs{\vec x - \vec x_0} \gtrapprox \abs{\mat \Sigma_0}$, the distance integration can be taken to infinity $\Delta \vec x \to \infty$.

\subsection{\sentence\NDRlong} \label{sec:ND distance integration}

Since the oscillation probability \eqref{eq:oscillation probability spacetime} depends on the absolute value square of the transition amplitude, the \EME \eqref{eq:NDR EME parameters} and \eqref{eq:NDR decay term parameters} need to be summed
\begin{align} \label{eq:NDR STEs and decay terms}
f_{ij} &= f_i + f_j = f_1 (\delta_i^2 + \delta_j^2) + \order*{\delta_i^4} \,, &
\gamma_{ij}(t) &= \gamma_i(t) + \gamma_j(t) = (\gamma_i + \gamma_j) t \,.
\end{align}
The same holds true for the \STE \eqref{eq:NDR STE}
\begin{equation} \label{eq:NDR STEs}
F_{ij}(t,\vec x) := F_i(t,\vec x) + F_j(t,\vec x)\,,
\end{equation}
Since dispersion effects are neglected in the \NDR, the \STE can be approximated by its \LO term in the mass splitting expansion
\begin{align} \label{eq:NDR STE LO}
F_{ij}(t, \vec x) &= F_0(t, \vec x) + \order*{\delta_i} \,, &
F_0(t, \vec x) &= \vec \Delta_0^\trans(t, \vec x) \mat \Sigma_0^{-1} \vec \Delta_0(t, \vec x) \,, &
\vec \Delta_0(t, \vec x) &= \vec v_0 t - \vec x \,.
\end{align}
Since it is quadratic in the spacial coordinates, there are no constant or linear terms.
The position of the minimum of the \STE is thus at \LO given by
\begin{align} \label{eq:NDR STE minimum}
\vec x_{ij}(t) &= \vec x_0(t) + \order*{\delta_i} \,, &
\vec x_0(t) &= \vec v_0 t \,.
\end{align}

\paragraph{Expansion}

The expansion of the \STE in $\vec x$ around $\vec x_{ij}(t)$ reads at \LO in the mass splitting expansion
\begin{equation} \label{eq:NDR STE expansion}
F_0(t,\vec x) = (\vec x - \vec x_0(t))^\trans \mat \Sigma_0^{-1} (\vec x - \vec x_0(t)) + \order*{\delta_i} \,.
\end{equation}
Since dispersion is neglected, there is no constant term.
The phase \eqref{eq:NDR phase expansion} is rewritten in terms of the deviation from the maximum in position space $\vec x - \vec x_{ij}(t)$ yielding
\begin{align}
\phi_i(t,\vec x) &= \phi_i(t) - \vec p_i \dot (\vec x - \vec x_{ij}(t)) \,, &
\phi_i(t) &= E_i t - \vec p_i \dot \vec x_{ij}(t) \,.
\end{align}
Since the oscillation probability \eqref{eq:oscillation probability spacetime} is proportional to the absolute value square of the transition amplitude, the phase difference
\begin{align}
\phi_{ij}(t,\vec x) &= \phi_i(t,\vec x) - \phi_j(t,\vec x) = \phi_{ij}(t) - \vec p_{ij} \dot (\vec x - \vec x_{ij}(t)) \,,
\end{align}
needs to be considered.
The constant term encodes the phase difference in time, while the linear coefficient consists of a momentum difference
\begin{align} \label{eq:NDR phase difference expansion}
\phi_{ij}(t) &:= \phi_i(t) - \phi_j(t) = E_{ij} t - \vec p_{ij} \dot \vec x_{ij}(t) \,, &
E_{ij} &= E_i - E_j \,, &
\vec p_{ij} &= \vec p_i - \vec p_j \,.
\end{align}
Using the mass splitting expansion of the energy \eqref{eq:NDR energy-mass splitting expansion} and momentum \eqref{eq:NDR mass splitting expansion} yields
\begin{align}
E_{ij} &= (E_0 - \vec v_0 \dot \vec p_1) \delta_{ij} + \order*{\delta_i^2} \,, &
\vec p_{ij} &= \vec p_1 \delta_{ij} + \order*{\delta_i^2} \,, &
\delta_{ij} &:= \delta_i - \delta_j \,,
\end{align}
and the appearing difference in the mass splitting expansion parameter \eqref{eq:NDR mass splitting expansion parameter} can be approximated to be
\begin{align} \label{eq:delta difference}
\delta_{ij} &= \frac{m_i^2 - m_j^2}{2 E_0^2} = \frac{m_{ij} m_0}{E_0^2} + \order*{\frac{m_{ij}^2}{E_0^2}} \,, &
m_{ij} = m_i - m_j \,.
\end{align}
It is used that the reconstructed mass $m_0$ cannot be far from the mean of the mass eigenstate masses $m$, since otherwise, it is not possible to have $\vec p_i \approx \vec p_0$ and $E_i \approx E_0$ for both mass eigenstates at the same time, and thus one of the two $f_i$ terms in the amplitude \eqref{eq:NDR transition amplitude} would lead to large damping of oscillations.
Therefore, the phase difference is at \LO in terms of the proper time
\begin{align} \label{eq:NDR phase difference}
\phi_{ij}(t) &= m_{ij} \tau(t) + \order*{\delta_i^2, \frac{m_{ij}^2}{E_0^2}}\,, &
\tau(t) &= \frac{m_0}{E_0} t \,.
\end{align}

\paragraph{Integration}

The integral \eqref{eq:oscillation probabillity distance average} can be evaluated as a Gaussian integral, using relation \eqref{eq:Gauss integral} with the linear term from the phase expansion \eqref{eq:NDR phase difference expansion} and the Hessian from the \STE expansion \eqref{eq:NDR STE expansion}
\begin{align}
\vec b &= \vec p_{ij} \,, &
\mat A &= 2 \mat \Sigma_0^{-1} \,,
\end{align}
resulting in the time-independent localisation term
\begin{equation} \label{eq:NDR localisation}
\Lambda_{ij} = \frac14 \vec p_{ij}^\trans \mat \Sigma_0 \vec p_{ij} + \order*{\delta_i^3} \,.
\end{equation}
The transition amplitude after this distance integration is, therefore,
\begin{align} \label{eq:NDR oscillation probability}
\mathcal P(t) &\propto \sum_{ij} \exp\left[- \lambda^\prime_{ij}(t) - \i \phi_{ij}(t)\right] \,, &
\lambda^\prime_{ij}(t) &= f_{ij} + \Lambda_{ij} + \gamma_{ij}(t) \,,
\end{align}
where the \STE and the decay term are given by \eqref{eq:NDR STEs and decay terms}, the phase is given by \eqref{eq:NDR phase difference}, and the localisation term is given by \eqref{eq:NDR localisation}.

\subsection{\sentence\TDRlong} \label{sec:TD distance integration}

The calculations for the \TDR are very similar to the ones described in \cref{sec:ND distance integration}.
However, compared to the \NDR dispersion is not neglected, and therefore, the approximation $v_i \approx v_j$ does not apply, which leads to a separation of wave packets of different mass eigenstates over time.
The sum of the \EME \eqref{eq:TDR EME parameters} and the decay terms \eqref{eq:TDR decay term parameters} are
\begin{align} \label{eq:TDR EMEs and decay terms}
f_{ij} &= f_i + f_j = f_1 (\delta_i^2 + \delta_j^2) + \order*{\delta_i^4}\,, &
\gamma_{ij}(t) &= \gamma_i(t) + \gamma_j(t) = (\gamma_i + \gamma_j) t \,,
\end{align}
The distance average is performed by extending the terms in the exponent of the transition amplitude \eqref{eq:TDR amplitude} around the minimum of the sum of the \STEs \eqref{eq:TDR STE} of the two mass eigenstates appearing in the oscillation probability
\begin{equation}
F_{ij}(t, \proj x) = F_i(t, \proj x) + F_j(t, \proj x) \,.
\end{equation}
The minimum is given at the position
\begin{equation}
\proj x_{ij}(t)
= \frac{\proj \Sigma_i(t) \proj v_j + \proj \Sigma_j(t) \proj v_i}{\proj \Sigma_i(t) + \proj \Sigma_j(t)} t
= \frac{\proj v_i + \proj v_j}{2} t + \order{\delta_i} \,,
\end{equation}

\paragraph{Expansion}

The expansion of the \STE yields
\begin{equation} \label{eq:TDR STE expansion}
F_{ij}(t, \proj x) = F_{ij}(t) + \frac12 Z_{ij} \left(\proj x - \proj x_{ij}(t)\right)^2 + \order[\big]{\abs*{\proj x - \proj x_{ij}(t)}^3} \,.
\end{equation}
where the constant term and the Hessian are
\begin{subequations}
\begin{align} \label{eq:TDR dispersion term}
F_{ij}(t) &= \frac{v_{ij}^2}{2} \frac{t^2}{\proj \Sigma_i(t) + \proj \Sigma_j(t)}
= \frac14 \proj \Sigma_0^{-1} \proj v_{ij}^2 t^2 + \order*{\delta_i^3} \,, &
v_{ij} = \proj v_i - \proj v_j \,, \\
Z_{ij}(t) &= \frac{1}{\proj \Sigma_i(t)} + \frac{1}{\proj \Sigma_j(t)}  = \frac{2}{\proj \Sigma_0(t)} + \order*{\delta_i} \,,
\end{align}
\end{subequations}
using the expansion \eqref{eq:TDR velocity-mass splitting expansion} the velocity difference can be written as
\begin{align}
v_{ij} &= \proj v_1 \delta_{ij} + \order*{\delta_i^2} = \frac{\proj p_1 - E_1 \proj v_0}{E_0} \delta_{ij} + \order*{\delta_i^2} \,,&
\delta_{ij} := \delta_i - \delta_j \,.
\end{align}
Since the expansion is performed around the minimum, the linear term is still absent.
However, since dispersion is not neglected, the constant term does not vanish.
The phase can be rewritten similarly to the \NDR in terms of a constant and a linear term
\begin{align}
\phi_i(t,\proj x) &= \phi_i(t) - \proj p_i (\proj x - \proj x_{ij}) \,, &
\phi_i(t) &= E_i t - \proj p_i \proj x_{ij} \,.
\end{align}
Since the phase contains the imaginary contributions to the amplitude, the difference between the two mass eigenstates appears in the oscillation probability
\begin{equation}
\phi_{ij}(t, \proj x) = \phi_i(t, \proj x) - \phi_j(t, \proj x) =
\phi_{ij}(t) + \proj p_{ij} (x - \proj x_{ij}) \,,
\end{equation}
the constant term and the linear coefficient are
\begin{align}
\phi_{ij}(t) &= E_{ij} t - \proj p_{ij} \proj x_{ij} \,, &
E_{ij} &= E_i - E_j \,, &
\proj p_{ij} &= \proj p_i - \proj p_j \,.
\end{align}
Using the expansions \eqref{eq:TDR mass splitting expansion,eq:TDR energy-mass splitting expansion} yielding
\begin{align}
E_{ij} &= E_1 \delta_{ij} + \order*{\delta_i^2} = (E_0 - \proj v_0 \proj p_1) \delta_{ij} + \order*{\delta_i^2} \,, &
\proj p_{ij}(t) &= \proj p_1 \delta_{ij} + \order*{\delta_i^2} \,,
\end{align}
together with the approximation \eqref{eq:delta difference}, leads for the phase difference to
\begin{align} \label{eq:TDR phase difference}
\phi_{ij}(t) &= m_{ij} \tau(t) + \order*{\delta_i^2, \frac{m_{ij}^2}{E_0^2}} \,, &
\tau(t) &= \frac{m_0}{E_0} t \,.
\end{align}
which is identical to the phase difference in the \NDR \eqref{eq:NDR phase difference}.

\paragraph{Integration of the oscillation probability}

The integration
\begin{equation}
\mathcal P(t) \propto \int[\d \proj x]_{\proj x_{ij} - \Delta \proj x}^{\proj x_{ij} + \Delta \proj x} \mathcal P(t, \proj x) \,,
\end{equation}
can be performed using \eqref{eq:Gauss integral} with
\begin{align}
b &= \proj p_{ij} \,, &
A &= \proj Z_{ij} \,,
\end{align}
resulting in the constant localisation term
\begin{equation} \label{eq:TDR localisation}
\Lambda_{ij} = \frac12 \proj Z_{ij}^{-1} p_{ij}^2 = \frac12 \frac{\proj \Sigma_i(t) \proj \Sigma_j(t)}{\proj \Sigma_i(t) + \proj \Sigma_j(t)} \proj p_{ij}^2 = \frac14  \proj \Sigma_0 \proj p_{ij}^2 + \order*{\delta_i^3} \,.
\end{equation}
Finally, the oscillation probability reads
\begin{align} \label{eq:TDR oscillation probability}
\mathcal P(t) &\propto \sum_{ij} \exp\left[\lambda^\prime_{ij}(t) - \i \phi_{ij}(t)\right] \,, &
\lambda^\prime_{ij}(t) &= f_{ij} + \Lambda_{ij} + F_{ij}(t) + \gamma_{ij}(t) \,,
\end{align}
where the \EME and the decay term are given by \eqref{eq:TDR EMEs and decay terms}, the phase difference is given by \eqref{eq:TDR phase difference}, the dispersion term is given by \eqref{eq:TDR dispersion term}, and the localisation term is given by \eqref{eq:TDR localisation}.

\section{Phase shift} \label{sec:phase correction}

For short times, corrections to the \LO expression of the phase \eqref{eq:oscillation}, which scale linearly with the decay width expansion parameter $\epsilon_i$ become relevant.
The expansion in the mass splitting expansion parameter $\delta_i$ is given in \eqref{eq:NDR phase difference,eq:TDR phase difference} in detail.
Therefore, we derive an analytical correction to the phase for parameter points in which $\epsilon_i$ is non-negligible and test how the numerical phase as obtained via the algorithm described in \cref{sec:pseudocode} deviates from the \LO expression for the phase \eqref{eq:oscillation}.

As presented in \cref{sec:JS}, the \NLO contribution in $\epsilon_i$ to the phase results in the usual decay term $\gamma_i(t, \vec p)$ \eqref{eq:time integrated phase and decay}.
However, there is also a contribution from the \EME, which yields
\begin{equation}
\phi_i(t, \vec x, \vec p) = E_i(\vec p) t - \vec p \dot \vec x - \gamma_i(\vec p) \left[\frac{e^{}_{iP}(\vec p)}{2 \sigma_{EP}^2} - \frac{e^{}_{iD}(\vec p)}{2 \sigma_{ED}^2} \right] + \order*{\epsilon_i^2} \,.
\end{equation}
While the direct contribution of the correction term to the phase $\phi_i(t, \vec x, \vec p)$ is negligible, it has a significant effect on the \STE \eqref{eq:NDR STE LO} since the corrections appear in the linear term
\begin{align}
\vec \Delta_0(t, \vec x) &= \vec v_0 t - \vec x + \gamma_0 \left(\frac{\vec u_P}{2\sigma_{EP}^2} - \frac{\vec u_D}{2\sigma_{ED}^2}\right) + \order*{\epsilon_i^2} \,.&
\gamma_0 &= \frac{m_0 \Gamma}{2E_0} \,,
\end{align}
This correction to the \STE results in a shift of the position of its minimum \eqref{eq:NDR STE minimum}, which is now at
\begin{equation}
\vec x_{ij}(t) = \vec v_0 t + \gamma_0 \left(\frac{\vec u_P}{2\sigma_{EP}^2} - \frac{\vec u_D}{2\sigma_{ED}^2}\right) + \order*{\delta_i, \epsilon_i^2} \,.
\end{equation}
The correction becomes relevant for \emph{large} decay widths since then the particle's lifetime becomes small, while simultaneously, the correction term becomes larger.
The resulting phase is then given by
\begin{equation}
\phi_{ij}(t) = \phi_{ij} + m_{ij} \tau(t) \,,
\end{equation}
where the constant phase shift is
\begin{equation} \label{eq:constant phase shift}
\phi_{ij} = - \vec p_{ij} \left(\frac{\vec u_P}{2\sigma_{EP}^2} - \frac{\vec u_D}{2\sigma_{ED}^2}\right) \gamma_0 \,.
\end{equation}
Since $\mat \Sigma_0$ is symmetric, this can be simplified to read
\begin{align}\label{eq:phase shift}
\phi_{ij} &= \frac{\vec u_D^\trans}{2\sigma_{ED}^2} \mat \Sigma_0^{-1} \frac{\vec u_D}{2\sigma_{ED}^2} m m_{ij} \epsilon_0 - (D\to P) + \order*{\epsilon_i^2, \delta_i^2, \frac{m_{ij}^2}{E_0^2}} \,, &
\epsilon_0 &= \frac{\gamma_0}{E_0} \,.
\end{align}

\begin{figure}
\begin{panels}{.49}
\includetikz{absphase}
\caption{Phase shift.} \label{fig:abs phase shift}
\panel{.51}
\includetikz{phasediff}
\caption{Relative phase shift.}  \label{fig:rel phase shift}
\end{panels}
\caption[Absolute value of the phase shift]{
The absolute value of the phase shift $\phi_{ij}$ in panel \subref{fig:abs phase shift} and the absolute value of the relative phase shift $\phi_{ij}^\text{rel}$, defined in \eqref{eq:rel phase shift}, in panel \subref{fig:rel phase shift} as functions of the decay width and mass splitting.
} \label{fig:phase shift}
\end{figure}

Numerical results for the phase shift are shown in \cref{fig:phase shift}.
As can be seen from \cref{fig:abs phase shift}, in the considered parameter region, the total value of the modulus of the phase shift is small for most of the parameter space except for part of the region with large $\Delta m$, where it can get large.
However, we remark that in this parameter region, many oscillations take place before the heavy neutrino decays, and thus the phase shift per oscillation is still small.
This is illustrated in \cref{fig:rel phase shift}, which shows the modulus of the relative phase shift
\begin{equation} \label{eq:rel phase shift}
\phi_{ij}^\text{rel} := \frac{\phi_{ij}}{\max(2\pi, \abs{m_{ij} \tau})}\,.
\end{equation}
Furthermore, comparing with \cref{fig:damping fixed m}, one can see that in the region with a large total phase shift, the damping $\lambda$ is very large, such that the oscillation term is strongly suppressed, making the phase shift practically unobservable.
In summary, the numerical results show that, for current collider simulations, one can safely neglect the phase shift in the considered parameter region.

However, outside the applicability region the phase shift can become significant, which can be seen from \eqref{eq:constant phase shift}.
The numerical results match this behaviour.
Since these results are based on values of the decay width larger than $\Gamma_{\JS}$, and since we do not see a physical reasons for the phase shift to become arbitrarily large, we neglect the phase shift for the main part of this work.

\begin{lstlisting}[
float,
captionpos=b,
caption={[Algorithm describing the strategy to calculate decoherence effects.]%
Algorithm describing the strategy to calculate decoherence effects.
},label=lst:code
]
define $E_0$, $\vec p_0$, $m_0$|\label{ln:external energy momenta}|
for $V$ in {$P$, $D$} do define $\sigma_{\vec xV}$, $\sigma_{\vec pV}$, $\sigma_{EV}$, $\Sigma_V$, $\vec v_V^{}$|\label{ln:widths}|
for $i$ in {4, 5} do define $m_i$, $\Gamma_i$|\label{ln:neutrino masses and widths}|

$t$ = rand.variate[expdistri[mean[$\Gamma_4$, $\Gamma_5$]]] $\flatfrac{E_0}{m_0}$|\label{ln:tof}|

for $V$ in {$P$, $D$} do
    $e_V[E,\vec p]$ = $E - E_0 - (\vec p - \vec p_0) \cdot \vec v_V^{}$
    $f_V[E, \vec p, \gamma]$ = $\flatfrac{\abs{\vec p - \vec p_0}^2}{(2 \sigma_{\vec pV})^2}$ + $(e_V^2[E, \vec p]$ - $e_V[0, \vec p] \flatfrac{\gamma^2}{E})/(2 \sigma_{EV})^2$|\label{ln:EME P}|

for $i$ in {4, 5} do
    $E_i[\vec p]$ = sqrt[$\abs{\vec p}^2 + m_i^2$]|\label{ln:mass es energy 4}|
    $\gamma_i[\vec p]$ = $\flatfrac{m_i \Gamma_i}{2 E_i[\vec p]}$|\label{ln:decay 4}|
    $f_i[\vec p]$ = $f_P[E_i[\vec p], \vec p, \gamma_i[\vec p]]$ + $f_D[E_i[\vec p], \vec p, \gamma_i[\vec p]]$ |\label{ln:EME 4}|
    $\lambda_i[\vec p]$ = $f_i[\vec p]$ + $\gamma_i[\vec p]t$
    $e_i[\vec p]$ = $\flatfrac{e_P[E_i[\vec p], \vec p]}{(2\sigma_{EP}^2)} - \flatfrac{e_D[E_i[\vec p], \vec p]}{(2\sigma_{ED}^2)}$
    $\phi_i[\vec p, \vec x]$ = $E_i[\vec p] t - \vec p \cdot \vec x$ - $\gamma_i[\vec p] e_i[\vec p]$|\label{ln:phase 4}|
    $\alpha_i[\vec p, \vec x]$ = $\lambda_i[\vec p]$ + $i \phi_i[\vec p, \vec x]$|\label{ln:exponential 4}|

for $i$ in {4, 5} do // momentum integral
    $\vec p_i$ = argmin[$\lambda_i[\vec p]$]|\label{ln:pmin 4}|
    $\mat \Sigma_i[\vec x]$ = hessian[$\alpha_i[\vec p, \vec x], \vec p$] at $\vec p = \vec p_i$|\label{ln:quad 4}|
    $\vec \Delta_i[\vec x]$ = $\partial_{\vec p}\phi_i[\vec p, \vec x]$ at $\vec p = \vec p_i$|\label{ln:lin 4}|
    $F_i[\vec x]$ = $\flatfrac{\vec \Delta_i[\vec x] \cdot \mat \Sigma_i^{-1}[\vec x] \cdot \vec \Delta_i[\vec x]}{2}$

$\alpha_{45}[\vec x]$ = $\alpha_4[\vec p_4, \vec x]$ + $F_4[\vec x]$ + conj[$\alpha_5[\vec p_5, \vec x] + F_5[\vec x]]$|\label{ln:STE}|

// distance integral
$\vec x_{45}$ = argmin[Re[$\alpha_{45}[\vec x]$]]|\label{ln:x0}|
$\mat Z_{45}$ = hessian[$\alpha_{45}[\vec x], \vec x$] at $\vec x = \vec x_{45}$|\label{ln:hess STE}|
$\vec \Pi_{45}$ = $\partial_{\vec x}$Im[$\alpha_{45}[\vec x]$] at $\vec x = \vec x_{45}$
$\Lambda_{45}$ = $\flatfrac{\vec \Pi_{45} \cdot \mat Z_{45}^{-1} \cdot \vec \Pi_{45}}{2}$|\label{ln:localisation}|

$\beta_{45}$ = $\Lambda_{45}$ + $\alpha_{45}[\vec x_{45}]$|\label{ln:final exponential}|
$\mathcal N_{45}$ = log[exp[$- 2 \lambda_4[\vec p_4]$]$/2$ + exp[$ - 2 \lambda_5[\vec p_5]$]$/2$]|\label{ln:normalisation}|

$\lambda_{45}$ = Re[$\beta_{45} + \mathcal N_{45}$]|\label{ln:damping}|
$\phi_{45}$ = Im[$\beta_{45} + \mathcal N_{45}$]|\label{ln:phase}|
\end{lstlisting}

\section{Numerical decoherence derivation} \label{sec:pseudocode}

In this section, the algorithm for the numerical computation of the damping parameter $\lambda$ is presented.
The algorithm expects the momenta of the external particles and the widths of the wave packets of the external particles as input.
Our estimates for the external widths in the process in \cref{fig:feynman diagram} are given in \cref{tab:widths}.
Realistic momentum configurations can be generated using a \MC generator such as \software{MadGraph} together with the implementation of the \pSPSS introduced in \cite{Antusch:2022ceb,FR:pSPSS}.

We present an algorithm for the derivation of the numerical results in \cref{lst:code}.
In the first lines, the kinematics of the event and the wave packet widths of the external particles are used to define the given quantities as input to the algorithm.
In particular, in \cref{ln:external energy momenta}, the kinematics of the external particles are used to define the reconstructed quantities \eqref{eq:reconstructed mass}, in \cref{ln:widths} the definitions \eqref{eq:total widths,eq:energy-momentum width,eq:width velocities} are used to define relevant widths, and in \cref{ln:neutrino masses and widths} the masses and decay widths of the heavy neutrino mass eigenstates are defined.
In \cref{ln:tof}, the propagation time between production and decay of the heavy neutrino superposition is drawn from an exponential distribution, defined by the mean decay width of the neutrinos.
In \cref{ln:EME P}, the \EMEs \eqref{eq:time integrated EME} at production and detection are defined, where the \LO corrections in the decay width expansion \eqref{eq:pole energy dependence}, are taken into account.
In the following lines, quantities for the mass eigenstates are defined.
In \cref{ln:decay 4}, the decay term is calculated,
\cref{ln:EME 4} defines the \EME for the heavy neutrino, and in
\cref{ln:phase 4} the phase, taking into account the imaginary part stemming from the decay width expansion of the \EME, is calculated.
All exponential terms relevant for the transition amplitude \eqref{eq:time integrated amplitude} are collected in \cref{ln:exponential 4}.
The integration over the three-momentum of the heavy neutrino wave packet is performed by approximating all terms up to second order around the maximum of the wave packet.
The maximum of the wave packet is defined in \cref{ln:pmin 4} by the minimum of the \EME, where the effects of the decay term are taken into account.
Subsequently, the terms of the exponential quadratic in $\vec p - \vec p_i$ are computed in \cref{ln:quad 4}.
Since the expansion is around the minimum of $f_i(\vec p) + \gamma_i(\vec p)$, only the complex phase has to be considered for the linear terms in \cref{ln:lin 4}.
The integration results in the \STE in \cref{ln:STE}, which is defined in \eqref{eq:NDR STE,eq:TDR STE}.
The following steps are valid for the $\mathcal A_i \mathcal A_j^*$ terms in the probability since the damping term is relevant for terms $i \neq j$, which are responsible for oscillations.
The distance integral is evaluated in the same fashion as the three-momentum integral.
The only terms in the exponential that depend on the distance are the \STE $F_{ij}(\vec x)$ and the complex phase $\phi_{ij}(\vec x, t)$.
The expansion is around the minimum of the \STE computed in \cref{ln:x0}.
Since the phase is linear in $\vec x$, the only contribution to the quadratic terms in $\vec x - \vec x_{ij}$ are given by the Hessian of the \STE computed in \cref{ln:hess STE}.
After the integration, the final exponent term is defined by
\begin{equation}
\mathcal A_i(t) \mathcal A_j^*(t) = \mathcal N_{ij}(t) \exp\left(-\beta_{ij}(t)\right)
\end{equation}
in \cref{ln:final exponential}.
The normalisation is computed in \cref{ln:normalisation} based on the condition \eqref{eq:normalisation}.
For the computation of the damping in the case of \NNOs in this paper, the decay terms $\gamma_i[\vec p_i]$ in $\beta$ and in $\mathcal N$ have been neglected, as we found them to be not significant.
Finally, the damping parameter and the complex phase are computed in \cref{ln:damping,ln:phase}.

While the algorithm is presented with the \NDR in mind, it can easily be applied to the \TDR by replacing vector quantities denoted with bold font by their projection onto the direction of~$\vec p_0$.

\dummyacronym{LDR}

\printbibliography

@article{Majorana:1937vz,
    author = "Majorana, Ettore",
    title = "{Teoria simmetrica dell\textquoteright{}elettrone e del positrone}",
    reportNumber = "RX-888",
    doi = "10.1007/BF02961314",
    journal = "Nuovo Cim.",
    volume = "14",
    pages = "171--184",
    year = "1937"
}

@article{Jacob:1961zz,
    author = "Jacob, R. and Sachs, R. G.",
    title = "{Mass and Lifetime of Unstable Particles}",
    doi = "10.1103/PhysRev.121.350",
    journal = "Phys. Rev.",
    volume = "121",
    pages = "350--356",
    year = "1961"
}

@article{Minkowski:1977sc,
    author = "Minkowski, Peter",
    title = "{$\mu \to e\gamma$ at a Rate of One Out of $10^9$ Muon Decays?}",
    reportNumber = "Print-77-0182 (BERN)",
    doi = "10.1016/0370-2693(77)90435-X",
    journal = "Phys. Lett. B",
    volume = "67",
    pages = "421--428",
    year = "1977"
}

@article{Gell-Mann:1979vob,
    author = "Gell-Mann, Murray and Ramond, Pierre and Slansky, Richard",
    title = "{Complex Spinors and Unified Theories}",
    eprint = "1306.4669",
    archivePrefix = "arXiv",
    primaryClass = "hep-th",
    reportNumber = "PRINT-80-0576",
    journal = "Conf. Proc. C",
    volume = "790927",
    pages = "315--321",
    year = "1979"
}

@article{Glashow:1979nm,
    author = "Glashow, S. L.",
    title = "{The Future of Elementary Particle Physics}",
    reportNumber = "HUTP-79-A059",
    doi = "10.1007/978-1-4684-7197-7_15",
    journal = "NATO Sci. Ser. B",
    volume = "61",
    pages = "687",
    year = "1980"
}

@article{Schechter:1980gr,
    author = "Schechter, J. and Valle, J. W. F.",
    title = "{Neutrino Masses in SU($2$)${}\times{}$U($1$) Theories}",
    reportNumber = "SU-4217-167, COO-3533-167",
    doi = "10.1103/PhysRevD.22.2227",
    journal = "Phys. Rev. D",
    volume = "22",
    pages = "2227",
    year = "1980"
}

@article{Mohapatra:1979ia,
    author = "Mohapatra, Rabindra N. and Senjanovic, Goran",
    title = "{Neutrino Mass and Spontaneous Parity Nonconservation}",
    reportNumber = "MDDP-TR-80-060, MDDP-PP-80-105, CCNY-HEP-79-10",
    doi = "10.1103/PhysRevLett.44.912",
    journal = "Phys. Rev. Lett.",
    volume = "44",
    pages = "912",
    year = "1980"
}

@article{Yanagida:1980xy,
    author = "Yanagida, Tsutomu",
    title = "{Horizontal Symmetry and Masses of Neutrinos}",
    reportNumber = "TU-80-208",
    doi = "10.1143/PTP.64.1103",
    journal = "Prog. Theor. Phys.",
    volume = "64",
    pages = "1103",
    year = "1980"
}

@article{Schechter:1981cv,
    author = "Schechter, J. and Valle, J. W. F.",
    title = "{Neutrino Decay and Spontaneous Violation of Lepton Number}",
    reportNumber = "SU-4217-203, COO-3533-203",
    doi = "10.1103/PhysRevD.25.774",
    journal = "Phys. Rev. D",
    volume = "25",
    pages = "774",
    year = "1982"
}

@article{Beuthe:2001rc,
    author = "Beuthe, Mikael",
    title = "{Oscillations of neutrinos and mesons in quantum field theory}",
    eprint = "hep-ph/0109119",
    archivePrefix = "arXiv",
    doi = "10.1016/S0370-1573(02)00538-0",
    journal = "Phys. Rept.",
    volume = "375",
    pages = "105--218",
    year = "2003"
}

@article{Kersten:2007vk,
    author = {Kersten, J\"orn and Smirnov, Alexei Yu.},
    title = "{Right-Handed Neutrinos at CERN LHC and the Mechanism of Neutrino Mass Generation}",
    eprint = "0705.3221",
    archivePrefix = "arXiv",
    primaryClass = "hep-ph",
    doi = "10.1103/PhysRevD.76.073005",
    journal = "Phys. Rev. D",
    volume = "76",
    pages = "073005",
    year = "2007"
}

@article{Alwall:2011uj,
    author = "Alwall, Johan and Herquet, Michel and Maltoni, Fabio and Mattelaer, Olivier and Stelzer, Tim",
    title = "{MadGraph 5}",
    subtitle = "{Going Beyond}",
    eprint = "1106.0522",
    archivePrefix = "arXiv",
    primaryClass = "hep-ph",
    reportNumber = "FERMILAB-PUB-11-448-T",
    doi = "10.1007/JHEP06(2011)128",
    journal = "JHEP",
    volume = "06",
    pages = "128",
    year = "2011"
}

@article{Alloul:2013bka,
    author = "Alloul, Adam and Christensen, Neil D. and Degrande, C\'eline and Duhr, Claude and Fuks, Benjamin",
    title = "{FeynRules 2.0}",
    subtitle = "{A complete toolbox for tree-level phenomenology}",
    eprint = "1310.1921",
    archivePrefix = "arXiv",
    primaryClass = "hep-ph",
    reportNumber = "CERN-PH-TH-2013-239, MCNET-13-14, IPPP-13-71, DCPT-13-142, PITT-PACC-1308",
    doi = "10.1016/j.cpc.2014.04.012",
    journal = "Comput. Phys. Commun.",
    volume = "185",
    pages = "2250--2300",
    year = "2014"
}

@article{Cvetic:2015ura,
    author = "Cvetic, Gorazd and Kim, C. S. and Kogerler, Reinhart and Zamora-Saa, Jilberto",
    title = "{Oscillation of heavy sterile neutrino in decay of $B \to \mu e \pi$}",
    eprint = "1505.04749",
    archivePrefix = "arXiv",
    primaryClass = "hep-ph",
    reportNumber = "USM-TH-335",
    doi = "10.1103/PhysRevD.92.013015",
    journal = "Phys. Rev. D",
    volume = "92",
    pages = "013015",
    year = "2015"
}

@article{ATLAS:2015gtp,
    author = "Aad, Georges and others",
    collaboration = "ATLAS",
    title = "{Search for heavy Majorana neutrinos with the ATLAS detector in $pp$ collisions at $ \sqrt{s} = 8$ TeV}",
    eprint = "1506.06020",
    archivePrefix = "arXiv",
    primaryClass = "hep-ex",
    reportNumber = "CERN-PH-EP-2015-070",
    doi = "10.1007/JHEP07(2015)162",
    journal = "JHEP",
    volume = "07",
    pages = "162",
    year = "2015"
}

@article{Anamiati:2016uxp,
    author = "Anamiati, G. and Hirsch, M. and Nardi, E.",
    title = "{Quasi-Dirac neutrinos at the LHC}",
    eprint = "1607.05641",
    archivePrefix = "arXiv",
    primaryClass = "hep-ph",
    reportNumber = "IFIC-16-48",
    doi = "10.1007/JHEP10(2016)010",
    journal = "JHEP",
    volume = "10",
    pages = "010",
    year = "2016"
}

@article{Apollinari:2017lan,
    author = "Aberle, O. and others",
    editor = "Apollinari, G. and Béjar Alonso, I. and Brüning, O. and Fessia, P. and Lamont, M. and Rossi, L. and Tavian, L.",
    title = "High-Luminosity Large Hadron Collider (HL-LHC)",
    subtitle = "Technical Design Report",
    version = "0.1",
    reportNumber = "CERN-2017-007-M",
    doi = "10.23731/CYRM-2017-004",
    journal = "CERN Yellow Reports",
    journalsubtitle = "Monographs",
    volume = "4",
    year = "2017"
}

@article{CMS:2018iaf,
    author = "Sirunyan, Albert M and others",
    collaboration = "CMS",
    title = "{Search for heavy neutral leptons in events with three charged leptons in proton-proton collisions at $\sqrt{s} = 13$ TeV}",
    eprint = "1802.02965",
    archivePrefix = "arXiv",
    primaryClass = "hep-ex",
    reportNumber = "CMS-EXO-17-012, CERN-EP-2018-006",
    doi = "10.1103/PhysRevLett.120.221801",
    journal = "Phys. Rev. Lett.",
    volume = "120",
    number = "22",
    pages = "221801",
    year = "2018"
}

@article{Antusch:2017ebe,
    author = "Antusch, Stefan and Cazzato, Eros and Fischer, Oliver",
    title = "{Resolvable heavy neutrino\textendash{}antineutrino oscillations at colliders}",
    eprint = "1709.03797",
    archivePrefix = "arXiv",
    primaryClass = "hep-ph",
    doi = "10.1142/S0217732319500615",
    journal = "Mod. Phys. Lett. A",
    volume = "34",
    number = "07n08",
    pages = "1950061",
    year = "2019"
}

@article{Drewes:2019fou,
    author = "Drewes, Marco and Hajer, Jan",
    title = "{Heavy Neutrinos in displaced vertex searches at the LHC and HL-LHC}",
    eprint = "1903.06100",
    archivePrefix = "arXiv",
    primaryClass = "hep-ph",
    reportNumber = "CP3-19-11",
    doi = "10.1007/JHEP02(2020)070",
    journal = "JHEP",
    volume = "02",
    pages = "070",
    year = "2020"
}

@article{Antusch:2019eiz,
    author = "Antusch, Stefan and Fischer, Oliver and Hammad, A.",
    title = "{Lepton-Trijet and Displaced Vertex Searches for Heavy Neutrinos at Future Electron-Proton Colliders}",
    eprint = "1908.02852",
    archivePrefix = "arXiv",
    primaryClass = "hep-ph",
    doi = "10.1007/JHEP03(2020)110",
    journal = "JHEP",
    volume = "03",
    pages = "110",
    year = "2020"
}

@article{ATLAS:2020xyo,
    author = "Aad, Georges and others",
    collaboration = "ATLAS",
    title = "{Search for long-lived, massive particles in events with a displaced vertex and a muon with large impact parameter in $pp$ collisions at $\sqrt{s} = 13$ TeV with the ATLAS detector}",
    eprint = "2003.11956",
    archivePrefix = "arXiv",
    primaryClass = "hep-ex",
    reportNumber = "CERN-EP-2019-219",
    doi = "10.1103/PhysRevD.102.032006",
    journal = "Phys. Rev. D",
    volume = "102",
    number = "3",
    pages = "032006",
    year = "2020"
}

@article{Esteban:2020cvm,
    author = "Esteban, Ivan and Gonzalez-Garcia, M. C. and Maltoni, Michele and Schwetz, Thomas and Zhou, Albert",
    title = "{The fate of hints: updated global analysis of three-flavor neutrino oscillations}",
    eprint = "2007.14792",
    archivePrefix = "arXiv",
    primaryClass = "hep-ph",
    reportNumber = "IFT-UAM/CSIC-112, YITP-SB-2020-21",
    doi = "10.1007/JHEP09(2020)178",
    journal = "JHEP",
    volume = "09",
    pages = "178",
    year = "2020",
    related = "NuFIT:2022",
    relatedstring = "Online:",
}

@article{Chrzaszcz:2020emg,
    author = "Chrz\k{a}szcz, Marcin and Drewes, Marco and Hajer, Jan",
    title = "{HECATE: A long-lived particle detector concept for the FCC-$ee$ or CEPC}",
    eprint = "2011.01005",
    archivePrefix = "arXiv",
    primaryClass = "hep-ph",
    reportNumber = "CP3-20-48",
    doi = "10.1140/epjc/s10052-021-09253-y",
    journal = "Eur. Phys. J. C",
    volume = "81",
    number = "6",
    pages = "546",
    year = "2021"
}

@article{Antusch:2020pnn,
    author = "Antusch, Stefan and Rosskopp, Johannes",
    title = "{Heavy Neutrino-Antineutrino Oscillations in Quantum Field Theory}",
    eprint = "2012.05763",
    archivePrefix = "arXiv",
    primaryClass = "hep-ph",
    doi = "10.1007/JHEP03(2021)170",
    journal = "JHEP",
    volume = "03",
    pages = "170",
    year = "2021"
}

@article{Blondel:2022qqo,
    author = "Blondel, A. and others",
    title = "{Searches for long-lived particles at the future FCC-$ee$}",
    eprint = "2203.05502",
    archivePrefix = "arXiv",
    primaryClass = "hep-ex",
    doi = "10.3389/fphy.2022.967881",
    journal = "Front. in Phys.",
    volume = "10",
    pages = "967881",
    year = "2022"
}

@online{FR:pSPSS,
    author = "Antusch, Stefan and Hajer, Jan and Rosskopp, Johannes",
    titleaddon = "FeynRules model file",
    title = "{pSPSS}",
    subtitle = "{Phenomenological symmetry protected seesaw scenario}",
    year = "2022",
    month = "10",
    url = "https://feynrules.irmp.ucl.ac.be/wiki/pSPSS",
    doi = "10.5281/zenodo.7268362",
    eprinttype = "github",
    eprintclass = "heavy-neutral-leptons",
    eprint = "pSPSS",
}

@article{Antusch:2022hhh,
    author = "Antusch, Stefan and Hajer, Jan and Rosskopp, Johannes",
    title = "{Beyond lepton number violation at the HL-LHC: Resolving heavy neutrino-antineutrino oscillations}",
    eprint = "2212.00562",
    archivePrefix = "arXiv",
    primaryClass = "hep-ph",
    month = "12",
    year = "2022"
}

@article{CMS:2022fut,
    author = "Tumasyan, Armen and others",
    collaboration = "CMS",
    title = "{Search for long-lived heavy neutral leptons with displaced vertices in proton-proton collisions at $ \sqrt{\mathrm{s}} = 13$~TeV}",
    eprint = "2201.05578",
    archivePrefix = "arXiv",
    primaryClass = "hep-ex",
    reportNumber = "CMS-EXO-20-009, CERN-EP-2021-264",
    doi = "10.1007/JHEP07(2022)081",
    journal = "JHEP",
    volume = "07",
    pages = "081",
    year = "2022"
}

@article{Antusch:2022ceb,
    author = "Antusch, Stefan and Hajer, Jan and Rosskopp, Johannes",
    title = "{Simulating lepton number violation induced by heavy neutrino-antineutrino oscillations at colliders}",
    eprint = "2210.10738",
    archivePrefix = "arXiv",
    primaryClass = "hep-ph",
    doi = "10.1007/JHEP03(2023)110",
    journal = "JHEP",
    volume = "03",
    pages = "110",
    year = "2023"
}

\end{document}